\documentclass[twocolumn]{aastex63}
\usepackage[table,xcdraw]{colortbl}
\usepackage{amsmath, amssymb}
\usepackage{longtable}

\newcommand{\uat}[2]{\href{http://astrothesaurus.org/uat/#2}{#1 (#2)}}

\begin{document}

\title{Functional Data Analysis for Extracting the Intrinsic Dimensionality of Spectra:\\Application to Chemical Homogeneity in the Open Cluster M67}

\author[0000-0002-7626-506X]{Aarya~A.~Patil}
\correspondingauthor{Aarya~A.~Patil}
\email{patil@astro.utoronto.ca}
\affiliation{David A. Dunlap Department of Astronomy \& Astrophysics, University of Toronto, 50 St George Street, Toronto ON M5S 3H4, Canada}
\affiliation{Dunlap Institute for Astronomy \& Astrophysics, University of Toronto, 50 St George Street, Toronto, ON M5S 3H4, Canada}

\author[0000-0001-6855-442X]{Jo~Bovy}
\affiliation{David A. Dunlap Department of Astronomy \& Astrophysics, University of Toronto, 50 St George Street, Toronto ON M5S 3H4, Canada}
\affiliation{Dunlap Institute for Astronomy \& Astrophysics, University of Toronto, 50 St George Street, Toronto, ON M5S 3H4, Canada}

\author[0000-0003-3734-8177]{Gwendolyn~Eadie}
\affiliation{David A. Dunlap Department of Astronomy \& Astrophysics, University of Toronto, 50 St George Street, Toronto ON M5S 3H4, Canada}
\affiliation{Department of Statistical Sciences, University of Toronto, 700 University Avenue, 9th Floor, Toronto, ON M5G 1Z5, Canada}

\author[0000-0002-0193-0993]{Sebastian~Jaimungal}
\affiliation{Department of Statistical Sciences, University of Toronto, 700 University Avenue, 9th Floor, Toronto, ON M5G 1Z5, Canada}

\received{2021 August 19}
\accepted{2021 November 1}
\begin{abstract}
High-resolution spectroscopic surveys of the Milky Way have entered the Big Data regime and have opened avenues for solving outstanding questions in Galactic archaeology. However, exploiting their full potential is limited by complex systematics, whose characterization has not received much attention in modern spectroscopic analyses. In this work, we present a novel method to disentangle the component of spectral data space intrinsic to the stars from that due to systematics. Using functional principal component analysis on a sample of $18,933$ giant spectra from APOGEE, we find that the intrinsic structure above the level of observational uncertainties requires ${\approx}$10 functional principal components (FPCs). Our FPCs can reduce the dimensionality of spectra, remove systematics, and impute masked wavelengths, thereby enabling accurate studies of stellar populations. To demonstrate the applicability of our FPCs, we use them to infer stellar parameters and abundances of 28 giants in the open cluster M67. We employ Sequential Neural Likelihood, a simulation-based Bayesian inference method that learns likelihood functions using neural density estimators, to incorporate non-Gaussian effects in spectral likelihoods. By hierarchically combining the inferred abundances, we limit the spread of the following elements in M67: $\mathrm{Fe} \lesssim 0.02$ dex; $\mathrm{C} \lesssim 0.03$ dex; $\mathrm{O}, \mathrm{Mg}, \mathrm{Si}, \mathrm{Ni} \lesssim 0.04$ dex; $\mathrm{Ca} \lesssim 0.05$ dex; $\mathrm{N}, \mathrm{Al} \lesssim 0.07$ dex (at 68\% confidence). Our constraints suggest a lack of self-pollution by core-collapse supernovae in M67, which has promising implications for the future of chemical tagging to understand the star formation history and dynamical evolution of the Milky Way.
\end{abstract}

\section{Introduction}

The physical processes that govern galaxy disk formation and evolution encode themselves in the structure of stellar populations. Discerning this structure places key constraints on models of galactic disks and provides data for cosmological simulations to replicate. In the Milky Way, we have the ability to resolve individual stars and characterize disk stellar populations using large samples \citep{rix_2013}. Modern wide-field spectroscopic surveys such as the Apache Point Observatory Galactic Evolution Experiment \citep[APOGEE,][]{majewski_2017}, GALAH \citep{de_silva_2015}, and Gaia-ESO \citep{gilmore_2012} provide medium- to high-resolution ($R \approx 10,000$) spectra of ${\sim}10^5$ stars over a wide range of Galactocentric distances ($3\,\mathrm{kpc} < R < 15\,\mathrm{kpc}$) and distances from the midplane ($\left| z\right| < 1\,\mathrm{kpc}$). Future surveys such as WEAVE \citep{dalton_2012}, MOONS \citep{cirasuolo_2014}, SDSS-V \citep{kollmeier_2017}, and 4MOST \citep{de_jong_2019} will increase this number to millions. Using these data, we can estimate radial velocities, stellar parameters, chemical abundances, and ages for a good portion of stars in the Galactic disk.

Accurate and precise stellar abundance estimates through high-resolution spectroscopy allow us to probe the detailed chemical and dynamical evolution of the Milky Way. We can understand this using a simplified view of Galactic evolution described as follows. Stars form in groups in well-mixed molecular clouds \citep{shu_1987, lada_2003, feng_2014}. They inherit the chemical composition of the star-forming cloud, and this birth chemistry evolves over a star’s lifetime in foreseeable ways. Eventually, mass loss and supernovae accompanying stellar death chemically enrich the interstellar medium. From a dynamical perspective, we know that several non-axisymmetric forces in the Galaxy change stellar orbits over time \citep{sellwood_2002}; this makes tracing the orbital history and birth location of stars through kinematics difficult. However, stellar abundances provide a fossil record of the star formation history, chemical enrichment, and dynamical evolution of the Galaxy. Therefore, they are of utmost importance to Galactic studies as well as to stellar astrophysics in general.

Despite orders-of-magnitude improvements in observations and significant advances in our understanding of stellar photospheres, stellar spectroscopy in its current form faces four major issues: (a) several instrumental/extrinsic factors affect observations, e.g., absorption by complex molecules in the interstellar medium \citep{zasowski_2015}, emission and absorption by the Earth's atmosphere, and persistence in the detector, (b) observations are high-dimensional -- $\mathcal{O}(10,000)$ wavelengths, (c) theoretical modeling of stellar photospheres and radiative transfer computations are expensive, and (d) current models are still incomplete, e.g., missing atomic and molecular data. The advent of Big Data in spectroscopy allows us to adopt data-driven approaches, but we still need to bridge the \textit{synthetic gap} between data and models to accurately and efficiently exploit its full potential \citep[e.g.,][]{obrian_2021}.

Several authors have adopted data-driven techniques to estimate stellar abundances \citep[e.g.][]{ness_2015, leung_2019}. While these improve the precision (not necessarily accuracy) of estimation, they still rely on high-quality model-derived abundances to construct training sets that incorporate domain-specific knowledge. Instead, we can directly characterize the \textit{intrinsic} variability of stellar spectra and retain information that models discard, remove degeneracies, and ensure accurate propagation of uncertainties. We then have the potential to investigate the chemical space spanned by a large number of stars, thereby understanding chemo-dynamical evolution on Galactic scales.

Working directly with the APOGEE spectral data, \cite{price_jones_2018} find the maximum number of dimensions needed to characterize multi-element abundance variations in spectra. They reduce the dimensionality of chemical space underlying APOGEE giants using expectation-maximized principal component analysis \citep[EMPCA;][]{dempster_1977, roweiss_1997} to ${\approx}10$ principal components (PCs). EMPCA adds an expectation-maximization likelihood step to PCA for handling noisy and missing data, an advantageous property for dealing with spectroscopic observations. However, their PCs have significant imprints of systematics, which illustrates that their empirical covariance is not the best estimate of the true, intrinsic population covariance.

In this paper, we apply functional principal component analysis (FPCA), a functional data analysis method \citep{ramsay_1997}, to assess the dimensionality of the intrinsic spectral space in the H-band, which is composed of stellar parametric and abundance information. FPCA can overcome some of the difficulties of classical PCA and EMPCA by adding continuity, a degree of smoothness, and domain knowledge to the computation of PCs. These features reduce the influence of noise and disentangle variance due to systematic errors. We find the intrinsic spectral structure to be ${\approx}10$-dimensional, which agrees with previous studies \citep{price_jones_2018, ting_2021}. However, our dimensions are dominated by absorption features, without any obvious systematics.

Beyond assessing the dimensionality of stellar spectra, our analysis has promising implications for the \textit{chemical tagging} of dispersed birth clusters within the Galaxy \citep{freeman_2002}. Chemical tagging finds groups of chemically similar stars in an attempt to trace them back to their birth locations. While this technique can revolutionize our understanding of Galactic evolution, it requires that birth clusters are chemically homogeneous, an assumption that is difficult to confirm observationally. \cite{bovy_2016} showed that using the spectra directly for studies of chemical homogeneity in open clusters provides stringent and reliable constraints compared to more traditional abundance-based studies. However, their method lacks the ability to interpret results -- e.g., one cannot determine the abundances of the open cluster in question. Instead, we can project spectra of stars in open clusters onto our systematic-free intrinsic structure, followed by extraction of chemical variation from that of stellar parameters using approaches like those in \cite{price_jones_2019} and \cite{mijolla_2021}. We can then constrain homogeneity in the resulting chemical space, whose abundance information is readily interpretable. 

To prove that the intrinsic structure defined by our functional PCs (FPCs) incorporates stellar photospheric information, we use it to infer stellar parameters and abundances of giants in the well-studied open cluster M67 and constrain its chemical homogeneity. We apply a simulation-based Bayesian inference method called Sequential Neural Likelihood to estimate complex spectral likelihoods and further bridge the synthetic gap. Our aim in this application is to (a) infer accurate and precise abundances using spectra projected onto our FPCs, (b) help validate chemical tagging by limiting the abundance spread in M67, and (c) provide evidence that we can directly \textit{tag} chemically similar groups in our intrinsic structure through, e.g., clustering algorithms like those in \cite{hogg_2016} and \cite{price_jones_2019} and forgo the use of model-derived abundances.

Our discussion leads us to the question: What is functional data analysis (FDA), and how do we treat stellar spectra as \textit{functional}? Spectroscopic observations span a range of wavelengths that are densely or sparsely distributed depending on the survey. Even the spectrum with the highest resolution and highest signal-to-noise ratio has an uncertainty associated with each flux and wavelength value. These noisy, discrete values reflect an underlying smooth curve of flux over wavelength or a spectral \textit{function} \citep{young_1994}. Therefore, we can treat spectral data as a sample of \textit{functional} observations, and this opens avenues to use FDA, a rich field of statistics that stems from influential work on growth curves \citep{rao_1958} and nonparametric regression \citep{eubank_1999}.

We can apply \textit{functional} counterparts of multivariate statistical methods to spectral analysis. FPCA is one such method that is well known in several scientific and industrial fields such as economics, finance, medicine, and meteorology \citep[][and references therein]{shahid_2013, wang_2016}. Within the last few years, this method has shown promising applications in different sub-fields of astronomy \citep[e.g.,][]{he_2018, weiler_2018, kou_2020, mandel_2020}. However, there are just a handful of these applications because astronomy has yet to leverage FPCA's key advantages. In particular, FDA is a powerful analysis framework for astronomical observations because they have a prototypical functional form dictated by the physical processes that generate them.

We describe the mathematical foundation of treating spectra as functional data in Section \ref{sec:functional} and use that to explain the FPCA algorithm. In Section \ref{sec:case}, we perform a case study on a sample of 10 spectra with 200 wavelength dimensions to illustrate the FPCA methodology and its advantages. Section \ref{sec:data} describes the APOGEE data we use for the applications in the remainder of the paper. In Section \ref{sec:dim_spec}, we reduce the dimensionality of our spectral sample using FPCA and discuss the results. We use our FPCs to determine the chemical homogeneity of open cluster M67 in Section \ref{sec:M67}. In Section \ref{sec:discussion}, we discuss the advantages, limitations, and potential improvements of our novel methods and motivate promising applications in Galactic archaeology as well as astronomy in general. We also discuss the implications of our results in this section. In the closing Section \ref{sec:conclusion}, we summarize the paper and discuss its main takeaways.

\section{Functional Data Analysis}\label{sec:functional}
Classical PCA is a commonly used method to find low-dimensional representations of \textit{discrete data} sets. Its main drawback is that it is sensitive to systematics, especially in cases where the sample sizes are small. To overcome these drawbacks, we use FPCA, a method that reduces the dimensionality of a set of \textit{continuous functions}.

Data in astronomy are often in the form of discrete observations over time, wavelength, or some other physical domain, and we can treat them as individual \textit{functions}. We describe such \textit{functional data} in the context of stellar spectroscopy in Section \ref{subsec:stellarfunction} and provide the mathematical procedure for converting spectral data to smooth functions in Section \ref{subsec:fun_register}. In Section \ref{subsec:FPCA}, we discuss the technical details of FPCA and how we can use it to reduce the dimensionality of spectra. For a more thorough mathematical treatment, see, e.g., \cite{ramsay_2006}.

\subsection{Stellar spectra as functional data}\label{subsec:stellarfunction}
Spectroscopic data are discrete observations of continuous functions of wavelength $f(\lambda)$ (or frequency) subject to noise. We can model them as
\begin{equation}\label{eq:func_form}
    y_{nm} = f_n(\lambda_m) + \epsilon_{nm}
\end{equation}
where the flux value $y_{nm}$ of the $n^\mathrm{th}$ observed spectrum at the wavelength $\lambda_m$ is a function $f_n$ sampled at the wavelength plus an error $\epsilon_{nm}$. In the statistics literature, the error $\epsilon_{nm}$ is often assumed to be independently and identically distributed (i.i.d.) with zero mean and constant variance to represent white noise. However, for spectroscopic data analysis, it is typically necessary to go beyond such random errors to account for nonstationarity and autocorrelations, especially in cases where there are systematic issues.

Spectral functions are bounded and square-integrable and therefore serve as \textit{points} in the $L^2$ Hilbert space, a generalization of Euclidean space to infinite dimensions \citep{debnath_2005}. The projections of such a function onto a set of orthogonal axes in the Hilbert space represent its coordinates; we can view it as a \textit{functional vector} and the spectral data corresponding to it as \textit{functional data}.

\subsection{Registering functional data as smooth functions}\label{subsec:fun_register}
The first step in FDA is the representation of data in a functional form. Our aim is to fit the discrete measurements $y_{nm},\,n=1,\dotsc,N;\,m=1,\dotsc,M$ to the model described in Equation~\eqref{eq:func_form} and filter out the noise $\epsilon_{nm}$. 
We express the functional form $f_n(\lambda)$ of the $n^\mathrm{th}$ spectrum as a linear combination of a collection or set of basis functions

\begin{equation}\label{eq:func_est}
    f_n(\lambda) = \sum_{k=1}^{K} \alpha_{nk} \phi_k(\lambda) = \boldsymbol{\phi}(\lambda)^\intercal \boldsymbol{\alpha}_n
\end{equation} where $\phi(\lambda)= \{\phi_1(\lambda),...,\phi_K(\lambda)\}$ is the set of $K$ basis functions that we can define as a $K \times M$ matrix $\boldsymbol{\phi}(\lambda)$ when viewed as a discrete set of observations and sampled at the same set of frequencies as that of the data; $\boldsymbol{\alpha}_n$ is a vector of length $K$ that contains the coefficients $\alpha_{nk}$. We estimate the coefficient vector $\boldsymbol{\alpha}_n$ by minimizing the sum of the squared deviations between the data and the functional representation

\begin{equation}\label{eq:SSE}
    \textsc{SSE}(\mathbf{y}_n \mid \boldsymbol{\alpha}_n)  = \sum_{m=1}^{M} \left[y_{nm} - \sum_{k=1}^{K} \alpha_{nk} \phi_k(\lambda_m) \right]^2
\end{equation} where SSE represents the sum of squared errors. We can express this in matrix format as
\begin{equation}\label{eq:SSE_vec}
    \textsc{SSE}(\mathbf{y}_n \mid \boldsymbol{\alpha}_n)  = \left( \mathbf{y}_n - \boldsymbol{\phi}^\intercal \boldsymbol{\alpha}_n \right)^\intercal (\mathbf{y}_n -  \boldsymbol{\phi}^\intercal \boldsymbol{\alpha}_n).
\end{equation} 

The simple least-squares approximation works for cases where the errors $\epsilon_{nm}$ are i.i.d. with $N(0, \sigma^2)$. In realistic situations, errors are heteroskedastic; as such we use weighted least-squares regressions by minimizing the criterion
\begin{equation}\label{eq:SSE_W_vec}
    \textsc{SSE}(\mathbf{y}_n \mid \boldsymbol{\alpha}_n)  = \left( \mathbf{y}_n -  \boldsymbol{\phi}^\intercal \boldsymbol{\alpha}_n \right)^\intercal \mathbf{W} (\mathbf{y}_n - \boldsymbol{\phi}^\intercal \boldsymbol{\alpha}_n)
\end{equation} where $\mathbf{W}$ is the weight matrix usually given by the inverse of the variance-covariance matrix $\boldsymbol{\Sigma}_e$ of reported uncertainties. There is also the possibility of penalizing the ordinary or weighted SSE to regulate the smoothness of the estimated function $\hat f_n(\lambda)$. In this approach, cross-validation techniques help choose the size of a roughness penalty. For our purposes, we find that the degree of smoothness provided by the least-squares fitting of basis functions is enough.

The basis expansion projects the infinite-dimensional random functions onto a finite-dimensional space. Generally, orthonormal basis functions such as Fourier series, Legendre polynomials, and B-splines form the basis, and the choice of basis functions depends on the nature of the data. However, due to the complex functional signatures of spectral data, the standard orthonormal bases are not appropriate. Basis expansions approximate functional data well when basis functions reflect the physical processes that generate the data. Therefore, we use model spectra as basis functions; model spectra are physically motivated, and their characteristics are similar to those underlying spectra. We explain this basis in more detail in Section \ref{subsec:case_FPCA}.

\subsection{Functional Principal Component Analysis}\label{subsec:FPCA}
Classical PCA aims to find orthogonal modes that incorporate maximal variation of a set of (multivariate) random vectors \citep{jolliffe_1986}. One can compute these modes and the variance they explain through eigendecomposition of the empirical covariance matrix. FPCA is the equivalent of doing this in the Hilbert space of random functions using the functional representations of data described in Section \ref{subsec:fun_register}.

Analogous to the covariance matrix in classical PCA, we define the covariance function
\begin{equation}
    v(\lambda_a, \lambda_b) = cov \left[ f(\lambda_a), f(\lambda_b) \right]
\end{equation} as the covariance between the wavelengths $\lambda_a$ and $\lambda_b$. Here $\lambda_a, \lambda_b \in [\lambda_\mathrm{start}, \lambda_\mathrm{end}]$, the wavelength range of the spectral data. The function $f$ is defined on all wavelengths in this range --- the range serves as its domain $\mathcal{D}$.

We can empirically approximate $v(\lambda_a, \lambda_b)$ using the functional representation of data $\hat f_n(\lambda)$ by
\begin{equation}\label{eq:cov}
    \hat v(\lambda_a, \lambda_b) = \frac{1}{N} \sum_{i=1}^{N} \left[\hat f_i(\lambda_a)\!-\!\hat\mu(\lambda_a)\right]\!\left[\hat f_i(\lambda_b)\!-\!\hat\mu(\lambda_b) \right]
\end{equation}
\begin{equation}
    \mathrm{where}\;\;\hat\mu(\lambda) =  \frac{1}{N} \sum_{i=1}^{N} \hat f_i(\lambda)
\end{equation} is the sample mean function that centers the data. 

We estimate the functional representations $\hat f_n(\lambda)$ of the observations $y_{nm}$ using Equation~\eqref{eq:SSE}. If $y_{nm}$ are centered, $\hat f_n(\lambda)$ will also be centered, and the resulting sample mean function $\hat\mu(\lambda) = 0$. We assume the data are centered, in which case Equation~\eqref{eq:cov} simplifies to

\begin{equation}\label{eq:cov_new}
    \hat v(\lambda_a, \lambda_b) = \frac{1}{N} {\boldsymbol{\phi}(\lambda_a)}^\intercal \boldsymbol{\alpha}^\intercal \boldsymbol{\alpha} \boldsymbol{\phi}(\lambda_b)
\end{equation} where $\boldsymbol{\alpha}$ is an $N \times K$ matrix representing the coefficients $\alpha_{nk}$ for $n=1,\dotsc,N$ observations.

Given the estimated covariance function $\hat v$, we can compute the PCs of functional data using Mercer's theorem \citep{mercer_1909}. Mercer's theorem states that if $v$ is continuous on the domain $\mathcal{D}^2$, then there exists an orthonormal set of continuous eigenfunctions $\{\Psi_j(\lambda), j=1,2,...\}$ with corresponding eigenvalues $\zeta_j$ such that

\begin{equation}\label{eq:mercer_eigen}
    (V \Psi_j)(\lambda) = \zeta_j \Psi_j(\lambda), \;\;\;\;\; \zeta_j \ge 0
\end{equation} where $V$ is the \textit{covariance operator} defined by the integral transform
\begin{equation}\label{eq:mercer_int}
    (V \Psi_j)(\lambda_a) = \int_{\mathcal{D}}^{} v(\lambda_a, \lambda_b) \Psi_j(\lambda_b) d\lambda_b.
\end{equation} 

While computing the eigenfunctions $\Psi_j(\lambda)$ from the covariance function $v$ seems to be an infinite-dimensional problem, the projection of data onto the finite collection of basis functions reduces it to a finite-dimensional one. These basis functions are said to \textit{span} the Hilbert space of spectral functions. Since the eigenfunctions belong to the same space, we may expand them onto the basis functions as follows:
\begin{equation}\label{eq:eigen_fun}
    \hat \Psi_j(\lambda) = \sum_{k=1}^{K} b_{jk} \phi_k(\lambda) = {\boldsymbol{\phi}(\lambda)}^\intercal \mathbf{b}_j
\end{equation} where $\mathbf{b}_j$ is a vector of length $K$ denoting the coefficients of the expansion.

Substituting Equations~\eqref{eq:cov_new} and \eqref{eq:eigen_fun} into Equation~\eqref{eq:mercer_int}, we obtain

\begin{equation}\label{eq:general}
\begin{split}
     (V \Psi_j)(\lambda_a) &= \int_{\mathcal{D}}^{} \hat v(\lambda_a, \lambda_b) \hat \Psi_j(\lambda_b)\,d\lambda_b\\
     &=\frac{1}{N} {\boldsymbol{\phi}(\lambda_a)}^\intercal \boldsymbol{\alpha}^\intercal \boldsymbol{\alpha} \mathbf{U} \mathbf{b}_j\\
\end{split}
\end{equation} where $\mathbf{U}$ is a $K \times K$ weight matrix whose entries are $U_{qr} = \int_\mathcal{D}^{} \phi_q(\lambda) \phi_r(\lambda) d\lambda \,\;\; q, r = 1,\dotsc,K$. We now substitute the eigenequation~\eqref{eq:mercer_eigen} into the left-hand side of Equation~\eqref{eq:general} to obtain
\begin{align}
\zeta_j \boldsymbol{\phi}(\lambda_a)^\intercal \mathbf{b}_j & = \frac{1}{N} \boldsymbol{\phi}(\lambda_a)^\intercal \boldsymbol{\alpha}^\intercal \boldsymbol{\alpha} \mathbf{U} \mathbf{b}_j, \;\;\; \forall \lambda_a \in \mathcal{D},\\
\rightarrow \quad \zeta_j \mathbf{b}_j &= \frac{1}{N} \boldsymbol{\alpha}^\intercal \boldsymbol{\alpha} \mathbf{U} \mathbf{b}_j.
\end{align}

This last set of equations is a finite-dimensional eigenproblem for $\zeta_j$ and $\mathbf{b}_j$, the eigenvalues and coefficients of the $j^{\mathrm{th}}$ eigenfunction respectively. Note that when the basis functions are orthonormal, $\boldsymbol{U}$ is the identity, and FPCA reduces to PCA on the basis coefficients $\boldsymbol{\alpha}$; however, in general, it differs.

\begin{deluxetable}{cl}
    \tablecolumns{2}
    \tablehead{
    \colhead{Symbol} &
    \colhead{Description}
    }
    \tablecaption{Mathematical Notation}
    \label{tab:notation}
    \startdata
        $\lambda$ & wavelength\\
        $n$ & index of spectrum of the $n^{\mathrm{th}}$ star\\
        $\mathcal{D}$ & wavelength domain of spectral data $[\lambda_\mathrm{start}, \lambda_\mathrm{end}]$\\
        $m$ & index of wavelength $\{\lambda_m\} \in \mathcal{D}$\\
        $N$ & number of spectra\\
        $M$ & number of wavelengths\\
        $y_{nm}$ & continuum-normalized flux measurement $f/f_c$\\
        $\mathbf{y}$ & $N \times M$ matrix of observations\\
        $\mathbf{W}$ & $M \times M$ weight matrix of each observation\\
        $\hat f_n(\lambda)$ & estimate of the functional form $f_n(\lambda)$ of $\mathbf{y}_n$\\
        $k$ & index of $k^{\mathrm{th}}$ basis function\\
        $K$ & number of basis functions\\
        $\phi(\lambda), \boldsymbol{\phi}$ & set of basis functions $\{\phi_k(\lambda)\}$, $K \times M$ matrix\\
        $\mathbf{U}$ & $K \times K$ weight matrix of basis functions\\
        $\alpha_{nk}, \boldsymbol{\alpha}$ & $k^{\mathrm{th}}$ basis expansion coefficient of $\mathbf{y}_n$, $N \times K$ matrix\\
        $v, V$ & covariance function, covariance operator\\
        $\hat \mu(\lambda)$ & sample mean function\\
        $j$ & index of eigenfunction\\
        $J$ & number of eigenfunctions used (dimensions)\\
        $\Psi(\lambda)$ & set of eigenfunctions $\{\Psi_j(\lambda)\}$\\
        $\zeta_j$ & eigenvalue of $\Psi_j(\lambda)$\\
        $b_{jk}, \mathbf{b}_j$ & basis expansion coefficients of $\Psi_j(\lambda)$, $K \times 1$ vector\\
        $sc_{nj}$ & score of $\mathbf{y}_n$ along $\Psi_j(\lambda)$\\
    \enddata
\end{deluxetable}

Armed with these eigenfunctions, we may project the data onto the first $J\le K$ components to obtain a lower-dimensional representation of the data -- the scores. The scores $sc_{nj},\,j=1,\dotsc,J$ of an observed spectrum $\mathbf{y}_n$, whose functional representation is $\hat f_n(\lambda)$, are as follows:

\begin{equation}\label{eq:scores}
    sc_{nj} = \int_{\mathcal{D}} \hat{f}_n(\lambda)\Psi_j(\lambda)\,d\lambda
\end{equation} which we can approximate using the Riemann sum
\begin{equation}\label{eq:scores_approx}
    sc_{nj} = \sum_{m=1}^{M-1} \hat f_n(\lambda_m) \Psi_j(\lambda_m)\,(\lambda_{m+1} - \lambda_m).
\end{equation} Essentially, $sc_{nj}$ quantifies the variability of the $n^{th}$ spectrum along the eigenfunction $\Psi_j$. Using these scores and the Karhunen–Loève theorem, we can obtain a dimensionally reduced approximation of $\mathbf{y}_{n}$ as
\begin{equation}\label{eq:dim_reduce}
    \hat y_{n}(\lambda) = \sum_{j=1}^{J} sc_{nj} \Psi_j(\lambda)
\end{equation} where $J$ represents the total number of scores (dimensions) used for the approximation.

In theory, we expect FPCA to provide better empirical estimates of covariance structures underlying data than PCA, especially in cases where the number of dimensions is larger than the number of observations. This expectation is because the \textit{functional representations of data} rather than the actual data compute the covariance structures. These representations are continuous, smooth, and physically motivated, making them less susceptible to noise and sample characteristics. To illustrate these advantages, we present a case study on a small sample of stellar spectra in the following section.

We refer the reader to Table~\ref{tab:notation} for an overview of the mathematical notation introduced in this section, which we continue to use in the remainder of the paper.

\section{Case study}\label{sec:case}
To illustrate the efficacy of FPCA for stellar-spectroscopic analysis, we consider the problem of computing the PCs of $N=10$ spectroscopic observations over $M=200$ wavelength points. Here $M$ is an order of magnitude larger than $N$; this helps us evaluate the merits of FPCA in cases where PCA runs into issues. In addition, we explore how well FPCA can reduce the influence of systematics by adding synthetic continuum-normalization errors to the sample. Section \ref{subsec:case_data} describes the spectral sample used for this case study as well as the simulation of continuum-normalization errors. Section \ref{subsec:case_FPCA} describes the computation of FPCs of the sample, and Section \ref{subsec:comp_PC} compares the resulting FPCs with the PCs obtained using classical PCA.

\begin{figure}[t]
    \centering
    \includegraphics[width=\linewidth]{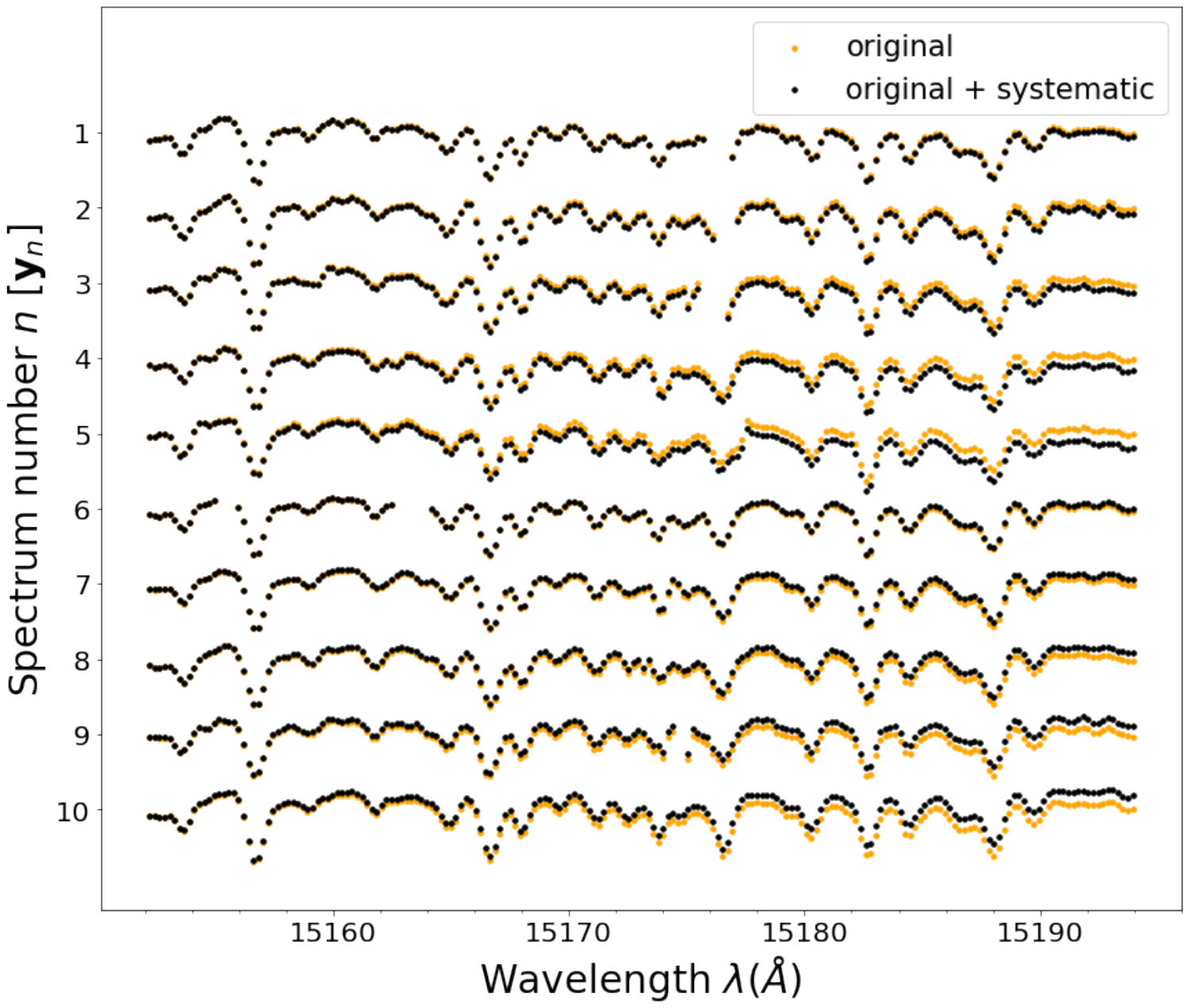}
    \caption{Stellar spectral samples used for the case study in Section \ref{sec:case}. The spectra labeled ``original" refer to the APOGEE subsample defined as $Y_{o}=\{\mathbf{y}_{o, 1}, \dotsc, \mathbf{y}_{o, 10}\}$, and those labeled ``original + systematic" have an added continuum-normalization systematic represented as $Y_{o+s}=\{\mathbf{y}_{o+s, 1}, \dotsc, \mathbf{y}_{o+s, 10}\}$ (refer to Section~\ref{subsec:case_data}). The two samples are continuum-normalized discrete observations of stellar spectra $y_{nm}$ where $n = 1, \dotsc, 10$ represents the spectrum number plotted on the y-axis, and $m = 1, \dotsc, 200$ is the wavelength index $\lambda_m$ for $\lambda_m \in [15152, 15195] \AA$ plotted on the x-axis. Note that the spectra are masked.}
    \label{fig:case_sample}
\end{figure}

\subsection{Data}\label{subsec:case_data}
For this case study, we use a subset of the APOGEE spectral sample that we describe in detail in Section \ref{subsec:sample}. We use estimates of effective temperature $T_\mathrm{eff}$ and surface gravity $\log g$ that are calibrated by the APOGEE Stellar Parameter and Chemical Abundances Pipeline (ASPCAP) to limit the APOGEE sample to those stars with $4270\,K < T_\mathrm{eff} < 4300\,K$ and $ 1.4 < \log g < 1.6$. We set these cutoffs around the stellar parameters of Arcturus, a well-studied red giant often used as a calibrator. The cuts result in a total of 20 spectra, and we choose 10 of these at random for this case study; we refer to them as $Y_{o} = \{\mathbf{y}_{o, 1},\dotsc, \mathbf{y}_{o, 10}\}$ or the ``original" sample. The remaining 10 spectra $B = \{\mathbf{b}_1,\dotsc, \mathbf{b}_{10}\}$ generate the basis functions, which we describe in detail in Section~\ref{subsec:case_FPCA}. Using the approach in \cite{ness_2015}, we continuum-normalize the $Y_{o}$ spectra (refer to Section~\ref{sec:data} for more details on the specific continuum normalization). In addition, we mask certain wavelengths in the spectra using the APOGEE \texttt{PIX\_MASK}, which flags erroneous pixels (Section \ref{subsec:apogee_error}). We defer detailed discussion of the data and its preprocessing until Section \ref{sec:data}, so that we may focus on illustrating FPCA with the case study.

To mimic incorrect continuum normalization, which is a systematic error commonly encountered in the analysis of APOGEE and other high-resolution spectra, we add a slope to the spectra, $Y_o$, of the order $\pm \, 0.00025 \, (\lambda - \lambda_\mathrm{start}) \, d$, where $d \in \mathbb{R}$ and $d \in [0, 5)$ represents the degree of the slope, and create a new sample of spectra: $Y_{o+s}$ or the ``original + systematic" sample. In this case study, we aim to evaluate whether FPCA has the ability to remove this continuum systematic. Note that we add and subtract the systematic in a manner such that the new and original sample mean spectra are equal; this ensures that the comparison of data variability between the two samples does not change in any essential way. Figure \ref{fig:case_sample} shows these samples.

\begin{figure}[t]
    \centering
    \includegraphics[width=\linewidth]{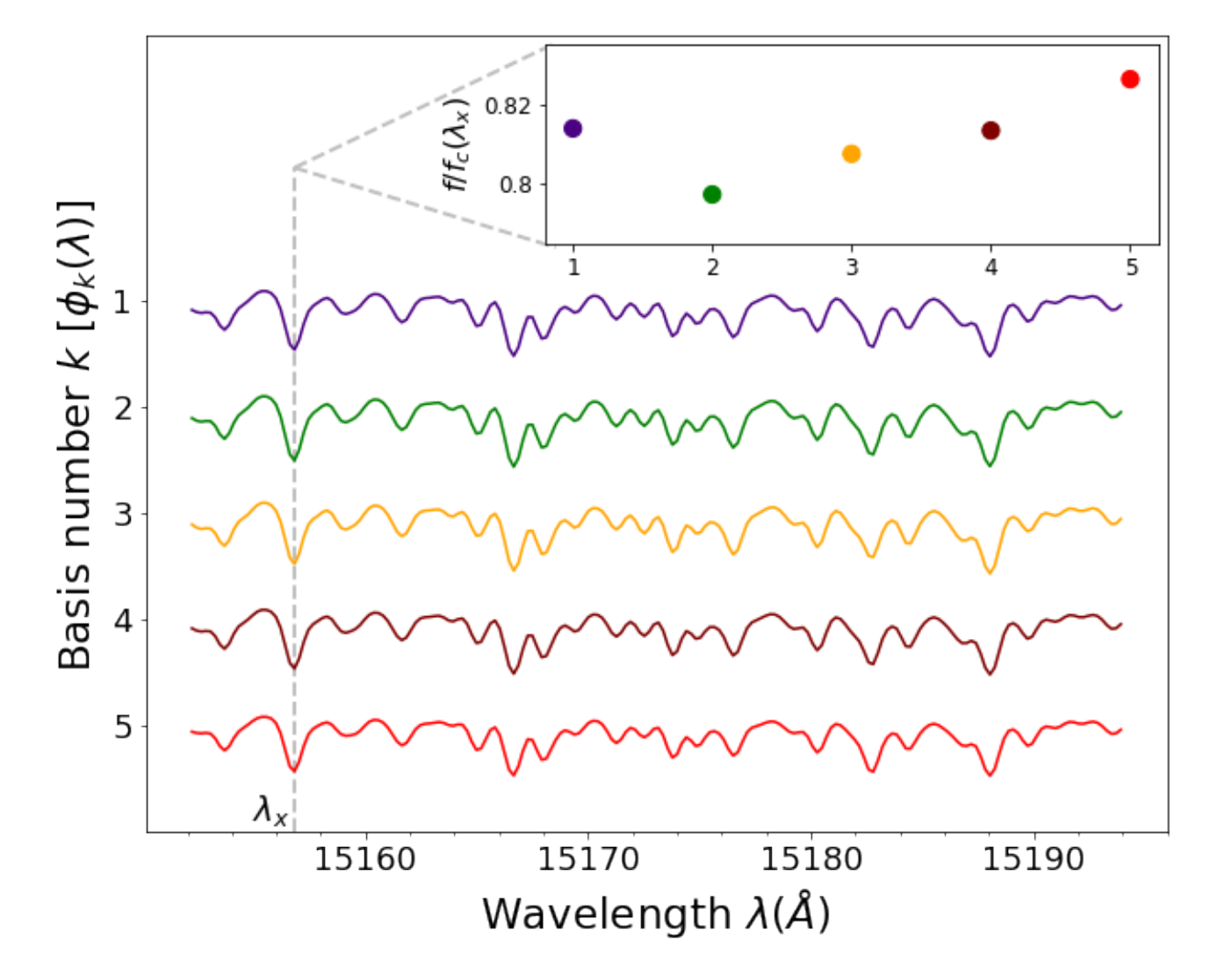}
    \caption{Model spectra used as basis functions in the case study (Section~\ref{subsec:case_FPCA}). Here $k$ represents the basis function number or index of $\phi_k(\lambda)$ where $k=1,\dotsc,5$. The wavelength range is given by $\lambda \in [15152, 15195] \AA$. Note that the model spectra are continuum-normalized. \textbf{Inset}: the flux variation ($f/f_c$) of the five basis functions at the wavelength $\lambda_x = 15156.82 \AA$. This variation illustrates that the basis functions mainly differ in their absorption lines due to variations in chemical abundances but overall look alike because of similar stellar parameters $T_{\mathrm{eff}}$ and $\log g$.}
    \label{fig:case_basis}
\end{figure}

\subsection{FPCA}\label{subsec:case_FPCA}
We now apply FPCA to explore the covariance structures of the two samples and reduce their dimensionality. The preliminary step is to register the spectra as smooth functions using basis expansion (refer to Section \ref{subsec:fun_register}). For this, we simulate model spectra using the polynomial spectral model (PSM) in \cite{rix_2016}. PSM emulates spectral synthesis with model atmospheres using a computationally efficient polynomial model. We use the ASPCAP-calibrated stellar parameter and abundance estimates of the spectral set $B = \{\mathbf{b}_1, \dotsc, \mathbf{b}_{10}\}$ (described in the previous section) as parameters to PSM; this generates the basis set

\begin{equation}\label{eq:basis_PSM}
    \phi_k(\lambda) = \textsc{PSM}(T_{\mathrm{eff}, k}, \log g_k, [\mathrm{X/H}]_k)
\end{equation} where $k=1,\dotsc,K$ represents the basis function number and $T_{\mathrm{eff}, k}, \log g_k$, and $[\mathrm{X/H}]_k$ are the stellar parameter and abundance estimates of spectrum $b_k$ for X = \{C, N, O, Na, Mg, Al, Si, S, K, Ca, Ti, V, Mn, Ni, Fe\}. Here we use a total of $K=5$ model spectra as basis functions; we selected these at random from the $10$. Section \ref{subsec:chooseK} provides the reasoning behind this choice of $K$. Figure \ref{fig:case_basis} shows the resulting basis functions.

\begin{figure}[t]
    \centering
    \includegraphics[width=\linewidth]{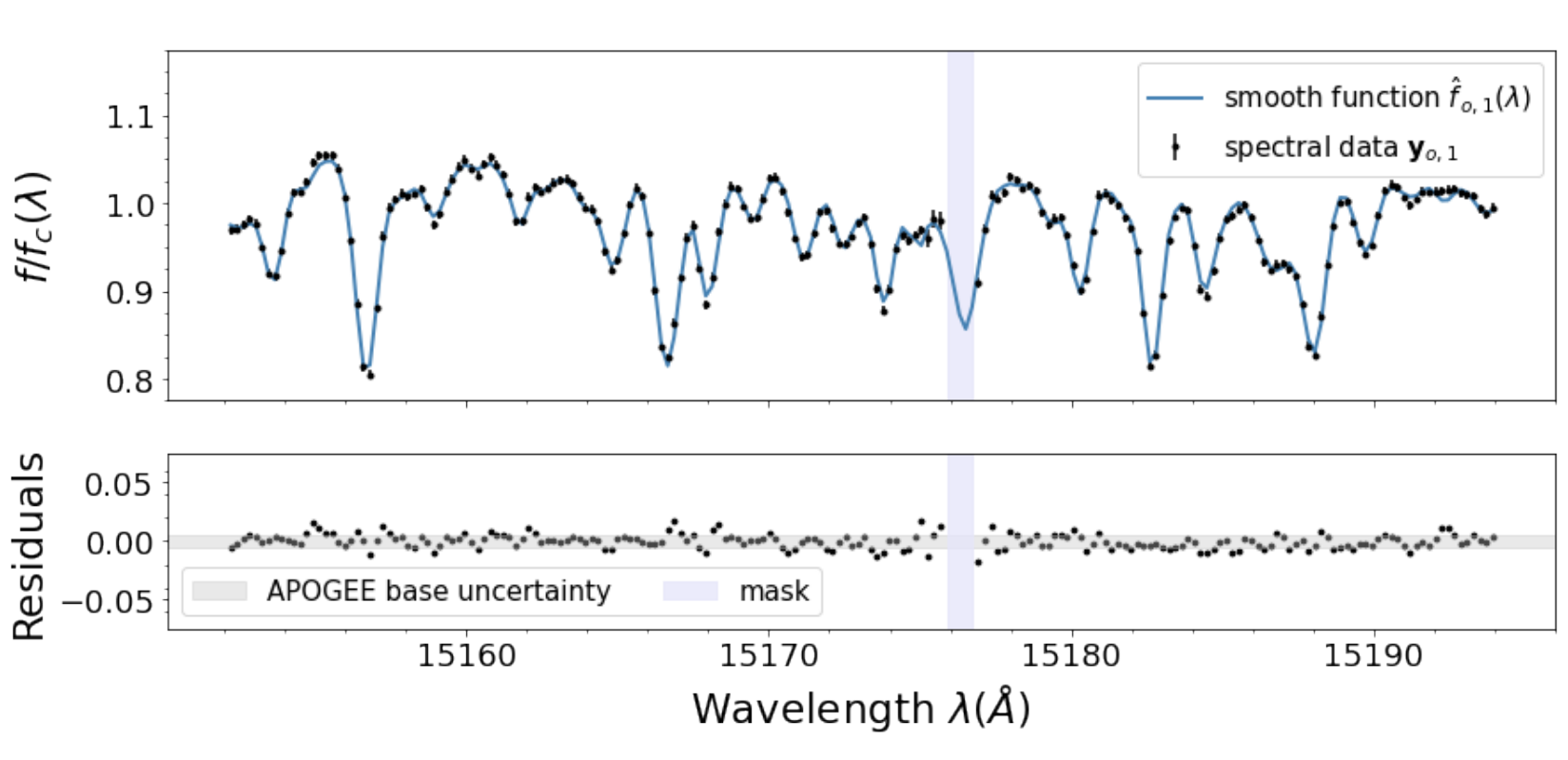}
    \caption{Functional representation of the continuum-normalized spectrum $\mathbf{y}_{o,1}$ in the ``original" sample of the case study (Section~\ref{subsec:case_FPCA}). \textbf{Top}: the approximated smooth function, $\hat f_{o,1}(\lambda)$ overplotted on the spectral data $\mathbf{y}_{o,1}$. \textbf{Bottom}: the residuals after subtracting the approximation from the data. The gray shaded horizontal region is from $[-0.005, 0.005]$ and signifies the APOGEE base uncertainty of 0.5\%. We use this base uncertainty to evaluate how well the approximation fits the data. The lavender horizontal region in both the panels shows that the functional approximation can generate missing (masked) data well.}
    \label{fig:case_func}
\end{figure}

The basis function expansion generates functional representations $\hat f_{o,n}(\lambda)$ and $\hat f_{o+s,n}(\lambda)$ of the masked spectra in the two samples $\mathbf{y}_{o,n}$ and $\mathbf{y}_{o+s,n}$ respectively. We use the weighted least-squares criterion (Equation~\eqref{eq:SSE_W_vec}) for estimating the coefficients $\boldsymbol{\alpha}_{o,n}$ and $\boldsymbol{\alpha}_{o+s,n}$ of the basis function expansion. APOGEE provides uncertainties corresponding to each spectral measurement, and the reciprocals of the squared uncertainties constitute the diagonal elements of the weight matrix $W$ in Equation~\eqref{eq:SSE_W_vec}. The resulting functional representations can generate data for the masked regions, a powerful feature of the basis expansion approach in FDA. Note that, in APOGEE data, the masked wavelengths differ from one spectrum to another, and basis expansion can deal with such variable masks. Figure \ref{fig:case_func} shows an example functional representation of a spectrum. The residuals of the data approximation are predominantly within the APOGEE base uncertainty, and they are unbiased, i.e., the mean error is zero. Thus, the approximation neither underfits nor overfits the data.

\begin{figure*}[ht!]
    \includegraphics[width=0.95\linewidth]{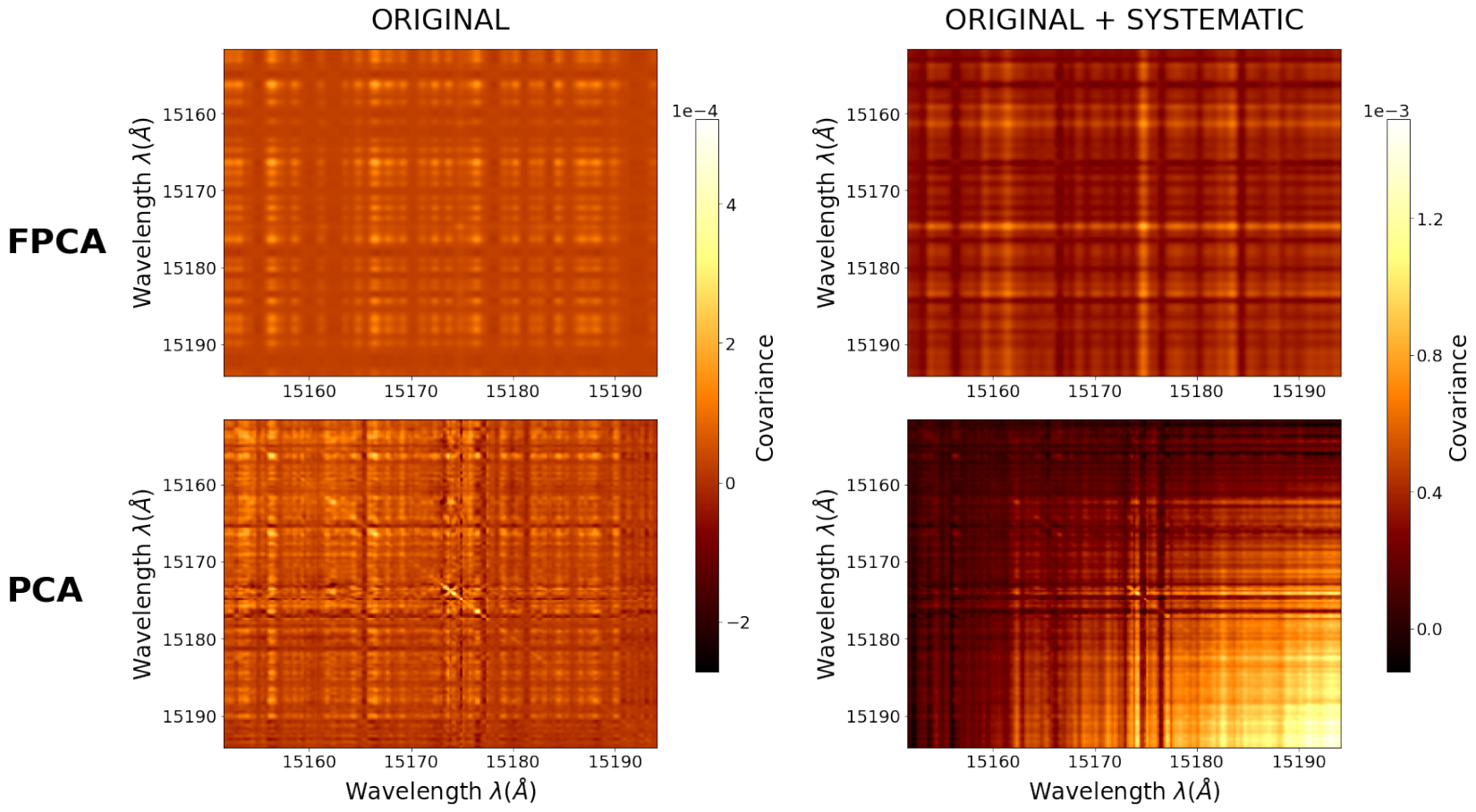}
    \caption{Comparison between FPCA and PCA covariance structures in the case study (Section \ref{subsec:comp_PC}). \textbf{Top left}: covariance function $v_o$ (refer to Equation~\eqref{eq:cov_new}) of the ``original" $Y_{o}$ sample. \textbf{Bottom left}: covariance matrix $\mathbf{C_o}$ of the $Y_{o}$ sample. \textbf{Right}:sSame information as the left panels, but for the ``original + systematic" sample ($v_{o+s}$, $\mathbf{C}_{o+s}$). We observe that the covariance functions are smooth, whereas the matrices are noisy -- $\mathbf{C}_{o+s}$ is significantly affected by the continuum-normalization systematic.
    } 
    \label{fig:cov}
\end{figure*}

\begin{figure*}[ht!]
    \includegraphics[width=0.95\linewidth]{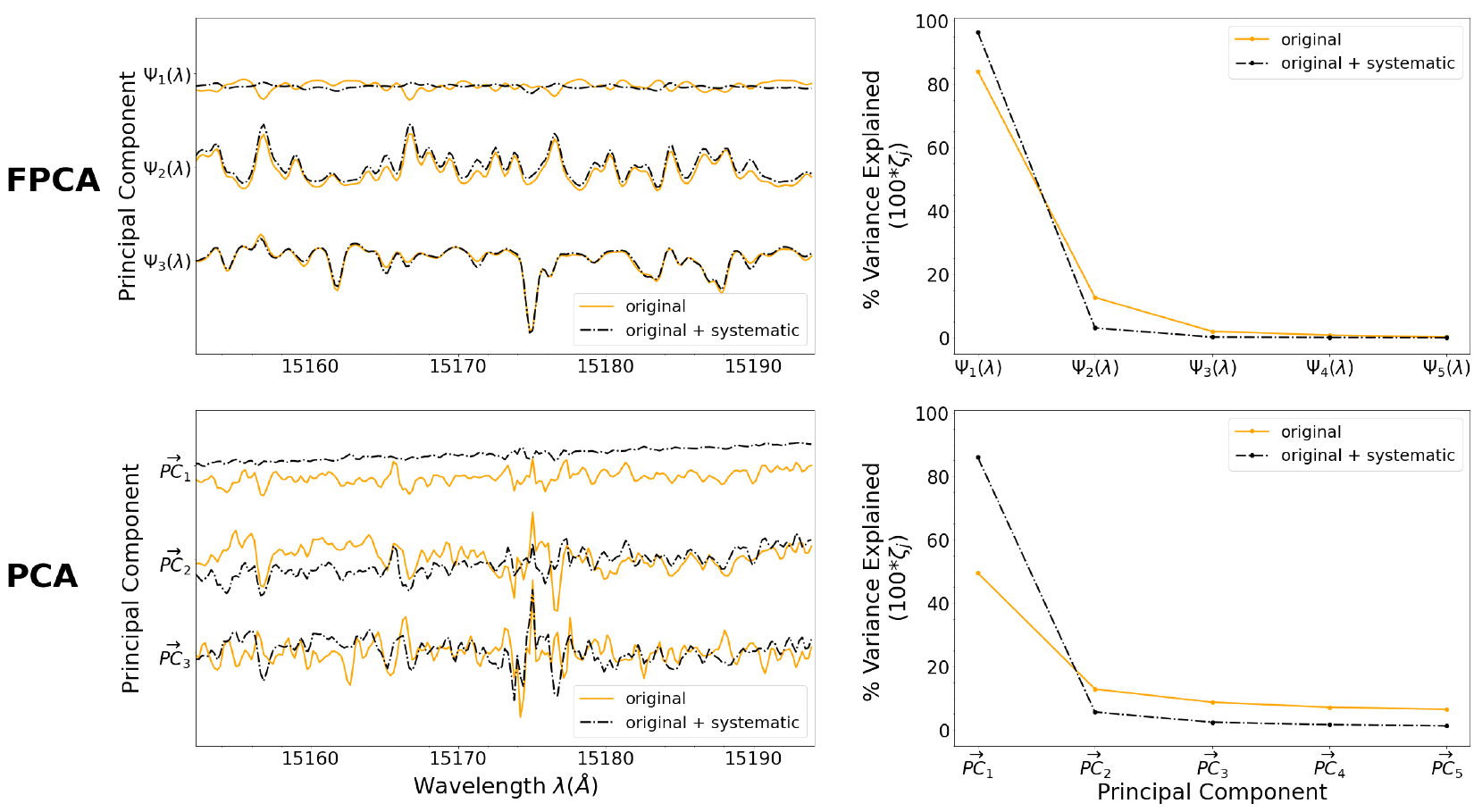}
    \caption{Comparison between FPCA and PCA eigenanalyses in the case study (Section~\ref{subsec:comp_PC}). \textbf{Top left}: first three FPCs $\Psi_{1,2,3}(\lambda)$ of the ``original" $Y_{o}$ and ``original + systematic" $Y_{o+s}$ samples. \textbf{Top right}: percentage variance explained by the FPCs of the two samples. \textbf{Bottom}: same information as the top panels, but for the classical PCs. It is clear that while noise and especially the continuum-normalization systematic strongly affect the classical PCA analysis, the FPCA analysis is unaffected by it.}
    \label{fig:case_pc}
\end{figure*}

We use the functional representations of the two samples to compute the corresponding covariance functions $v_{o}$ and $v_{o+s}$, which we show in the top panels of Figure \ref{fig:cov}. The top left panel of Figure \ref{fig:case_pc} displays the resulting FPCs or eigenfunctions $\Psi_j(\lambda)$, whereas the top right panel shows the percentage of total variance explained by them, i.e., $100\, \zeta_j$ where $\zeta_j$ are the eigenvalues. In the multivariate statistics literature, this is referred to as a \textit{scree} plot: a line plot that displays the eigenvalues of principal components.


\subsection{FPCA versus PCA}\label{subsec:comp_PC}
We apply classical PCA to the two spectral samples, ``original" $Y_{o}$ and ``original + systematic" $Y_{o+s}$, shown in Figure \ref{fig:case_sample}, and compare its results with those obtained in the previous section using FPCA. To ensure that a fair comparison is made, we impute masked values in the two samples with the sample mean spectrum, average the imputed data, and then apply classical PCA.

We show the covariance matrices of the two samples $\mathbf{C}_{o}$ and $\mathbf{C}_{o+s}$ in the bottom panels of Figure \ref{fig:cov}, and the eigenanalysis results in the bottom panels of Figure \ref{fig:case_pc}. Figures \ref{fig:cov} and \ref{fig:case_pc} helps us conclude the following:

\begin{enumerate}
    \item FPCs for both spectral samples are smoother than the noisy classical PCs and predominantly show absorption-like features; this is because of the assumptions on continuity and smoothness of the functional representations, which are inherited from the basis being physics-driven.
    \item FPCs of the $Y_{o}$ sample pick out similar variations to those of the $Y_{o+s}$ sample, and this indicates that the addition of a continuum-normalization systematic does not affect the FPCs. In contrast, the first classical PC of the $Y_{o+s}$ sample has a slope of $m \approx 0.0025$, which represents the systematic. We do not observe such a slope in the FPCs because our basis functions, being model spectra, only allow physically expected variations. One could, in principle, include the continuum-normalization variance in the FPCs by using basis functions such as the first few Legendre polynomials along with model spectra; however, one should use those basis functions that incorporate signal variability in data, not noise.
    \item The first FPC of the $Y_{o}$ sample explains most of the variance (${\sim}$85 \%) in the spectral data. In contrast, we need at least a few classical PCs to incorporate the same amount of variability in the sample. Note that this is a consequence of the fact that the eigenvalues obtained using FPCA are fundamentally different from those obtained using classical PCA. In the classical case, eigenvalues are computed using eigendecomposition of the covariance matrix, which quantifies the total variance in the data: ``intrinsic + noise."  In the functional case, eigenvalues are computed using Mercer's theorem on the covariance function, which quantifies variance in the functional representations of data (refer to Equation~\eqref{eq:mercer_eigen}); these eigenvalues indicate the contribution of FPCs in explaining intrinsic variance as opposed to any noise identified by the functional procedure.
    \item Both the FPCA and PCA explained-variance plots (scree plots) for the $Y_{o+s}$ sample show a significant first component because the continuum is the dominant source of variation in this sample.
    \item A direct interpretation of the covariance function of functional data or the covariance matrix of multivariate data is difficult. Instead, we use principal components to describe the data variability more intuitively. However, it is still informative to look at the covariance structures to understand the salient differences between FPCA and PCA. In Figure \ref{fig:cov}, we see that the top panels displaying $v_{o}$ and $v_{o+s}$ are smoother than the corresponding matrices $\mathbf{C}_{o}$ and $\mathbf{C}_{o+s}$, in the bottom panels; the matrices have features such as noisy backgrounds, outliers, and influence of systematic effects. Specific examples of these are: (1) $\mathbf{C}_{o}$ in the bottom left panel is overall noisier than $v_{o}$ in the top left panel and has some outliers around 15,175 $\AA$. (2) $\mathbf{C}_{o+s}$ in the bottom right panel is dominated by a gradient from short to long wavelengths due to the added continuum-normalization systematic, whereas $v_{o+s}$ in the top right panel is unaffected by it. These differences illustrate the advantages of FPCA over classical PCA.
\end{enumerate}

Our case study illustrates the advantages of using FPCA over PCA. It is evident that PCA is sensitive to noise, especially in cases where $N \ll M$, whereas FPCA attempts to remove all noise in the data without having access to an accurate noise model. This observation agrees with theoretical expectations and case studies performed across different fields in science and engineering \citep[e.g.,][]{viviani_2005, ramsay_2007}. The advantages of FPCA are also applicable in comparison with EMPCA and will be investigated further in Section \ref{sec:dim_spec}.

\subsection{How to choose $K$ and $J$?}\label{subsec:chooseK}
We use $K=5$ basis functions in the case study for illustrative purposes. In theory, $K$ can be any integer $ \in \{2,\dotsc,M\}$. However, the choice is dependent on the trade-off between bias and variance. As $K$ increases, the bias in the estimate of the functional form reduces, whereas the variance of the estimate increases. The latter may introduce noise in the estimate. We tested this by incrementally increasing $K$ for the ``original + systematic" $Y_{o+s}$ sample, and evaluating the resulting FPCs. We observed that the slope of the first FPC increased with increasing $K$, and this indicated a greater influence of the continuum-normalization systematic on the FPCs. Based on this qualitative assessment, we set $K = 5$ for the $Y_{o+s}$ sample.

\begin{figure}[t]
    \centering
    \includegraphics[width=\linewidth]{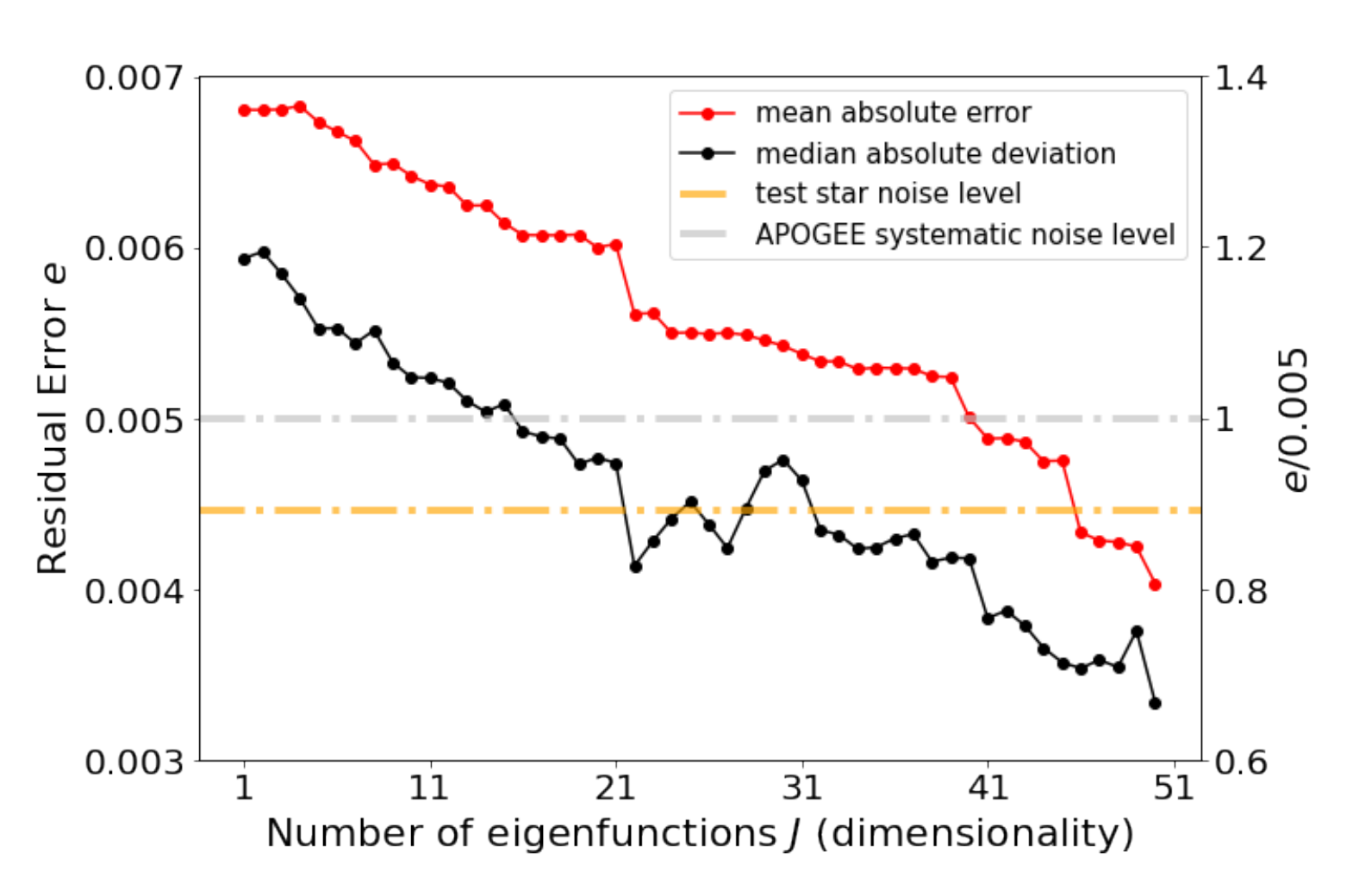}
    \caption{Mean absolute error (red) and median absolute deviation (black) of the test star model as a function of increasing $J$, the number of FPCs. The dashed orange line indicates the test star uncertainty level, which is the inverse of its signal-to-noise ratio, whereas the gray dashed line at $0.5\%$ indicates the base APOGEE systematic uncertainty. The y-axis on the left represents the absolute residual error $e$, whereas the one on the right, scaled as $e/0.005$, is in units of the base APOGEE uncertainty.
    }
    \label{fig:mse_validation}
\end{figure}

For data samples with a high signal-to-noise ratio, like the ``original" $Y_{o}$ sample, we can choose a larger $K$ to better fit the data and reduce the bias in the estimates of functional form. This is because the risk of fitting noise in the data as the variance of the estimate increases is low due to the data quality. In this case, $K$ should be large enough to incorporate all the signal information in the functional fits but not so large that the residuals are far smaller than the level of uncertainty. We observe this in Figure \ref{fig:case_func}, where the residuals trace the uncertainty well. While our qualitative analysis allows us to choose a decent value of $K$, it is desirable to have a more quantitative analysis. In the FDA literature, one can find several algorithms for choosing $K$ \citep{ramsay_2006}. However, there is no standard methodology, and this is potentially due to the discrete nature of $K$ \citep{ramsay_2006}. We use the \textit{stepwise variable selection} approach, which incrementally adds basis functions until some criterion is fulfilled; in our case, we add functions until the residuals of the fits have standard deviation within the base uncertainty. Using this approach, we find that $K \approx 50$ is best suited for the $Y_{o}$ sample. We note that while choosing $K$ is still an open question, domain knowledge can often assist in the selection.

After choosing the \textit{best} $K$ and computing FPCs, the next important question is to choose the number of FPCs $J$ that we use to perform dimensional reduction of the data. The value of $K$ sets the maximum $J$, and so $J \in \{1, \dotsc, K\}$.

We estimate the optimal $J$ for the $Y_{o}$ sample by modeling a test star that is similar to the sample, but not a part of it, using the FPCs. Essentially, we treat the sample as the training set and the test star as the validation set. We use Equation~\eqref{eq:dim_reduce} to project the test star onto the FPCs $\Psi_j$ and obtain the scores $sc_{nj}$. The projection of the test star onto the FPCs is done by applying the expectation-maximization (EM) algorithm \citep[][]{dempster_1977}, which ensures that masked and noisy regions of the test star are properly downweighted. We iteratively increase $J$ and compare the mean absolute error and median absolute deviation of the test star model to the noise level of the star as well as the base APOGEE systematic noise level. Figure \ref{fig:mse_validation} shows this comparison and illustrates that choosing $J$ is not straightforward; we find that $J \approx 10$ ensures that systematics do not affect the FPCs, whereas $J \approx 40$ counters measurement noise. The choice depends on the application.

\begin{figure}[t]
    \centering
    \includegraphics[width=0.97\linewidth]{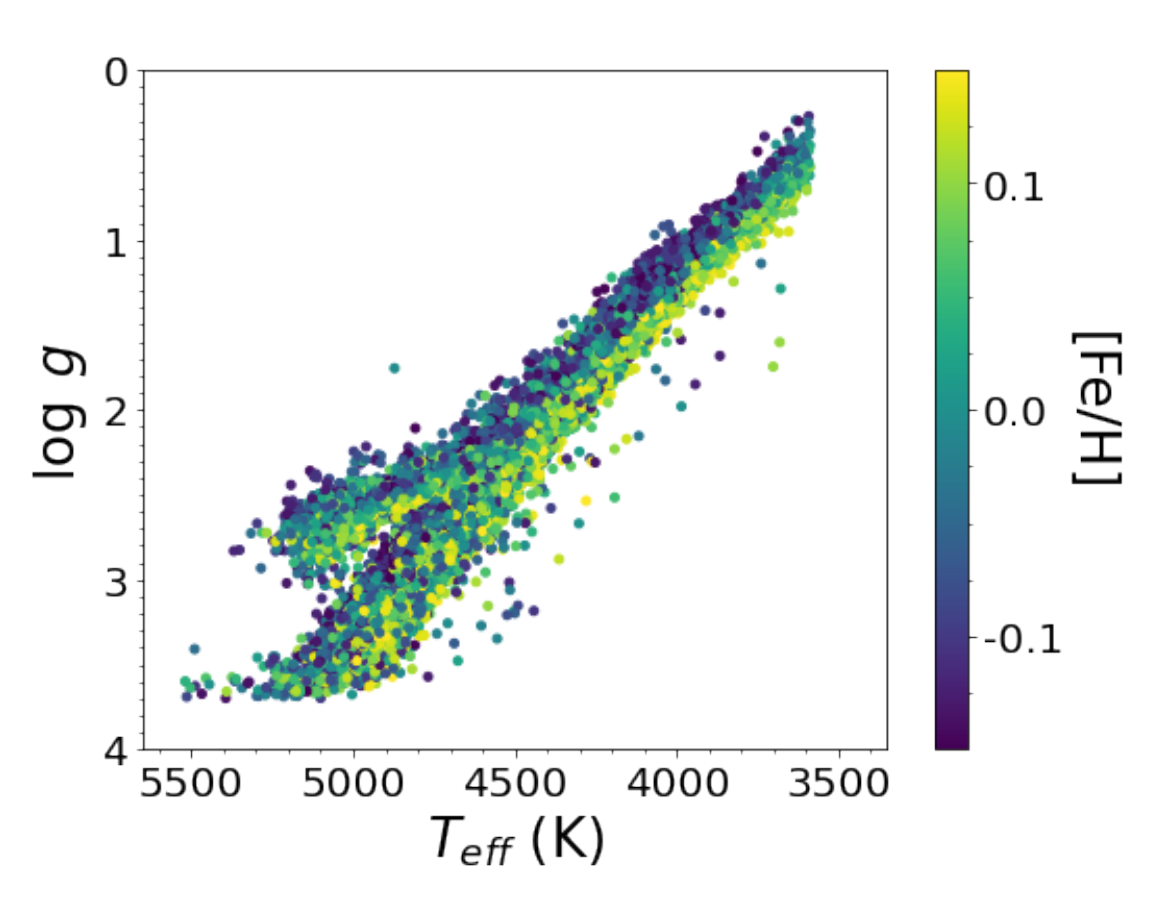}
    \caption{Spectroscopic H-R diagram ($\log g$ vs. $T_\mathrm{eff}$) of the APOGEE sample of $18,933$ giants used in this analysis. Each point is color-coded based on its metallicity $[$Fe/H$]$. The $T_{\mathrm{eff}}$, $\log g$, and $[$Fe/H$]$ estimates are ASPCAP-calibrated fits as described in Section \ref{sec:data}.}
    \label{fig:sample}
\end{figure}

\section{Data}\label{sec:data}
We use spectroscopic data from APOGEE \citep[][]{majewski_2017}, part of the Sloan Digital Sky Survey IV \citep[SDSS-IV;][]{blanton_2017}, for this work. The APOGEE spectroscopic survey provides high-resolution (R $\sim$ 22,500) near-infrared (H-band $1.51-1.69 \mu \mathrm{m}$) spectra for stars belonging to all stellar populations of the Milky Way - Galactic bulge, disk, and halo. The APOGEE multi-fiber spectrographs \citep{wilson_2019} can observe 300 targets simultaneously and are mounted on 2.5 m telescopes at Apache Point Observatory \citep{gunn_2006} and Las Campanas Observatory \citep{bowen_1973}. We use data from the public Data Release 14 \citep[DR14;][]{abolfathi_2018} of SDSS, which includes spectra for over 250,000 stars \citep{holtzman_2018, jonsson_2018}.

APOGEE provides repeat observations or ``visits" of most stellar targets in the DR14 sample. These are corrected for radial velocity, combined into one spectrum using ``global" or ``local" weighting, and supplied along with the individual visits.

ASPCAP \citep[][]{garcia_perez_2016} provides estimates of stellar parameters and chemical abundances for these combined spectra. The pipeline follows a two-step process. First, it computes a grid of synthetic spectra using custom linelists \citep{shetrone_2015} and finds the best-fitting spectrum for each combined APOGEE spectrum using \texttt{FERRE} \citep{allende_2006}. \texttt{FERRE} is capable of interpolating the spectral grid, and uses a $\chi^2$ minimization approach for fitting. Note that both the synthetic and combined APOGEE spectra are pseudo-continuum-normalized before making any comparison. The best fits result in estimates of atmospheric parameters: $T_{\mathrm{eff}}$, $\log g$, $\mathrm{vmicro}$, $\mathrm{vmacro}$, $\mathrm{vsini}$, and $[\mathrm{M/H}]$, $[\alpha\mathrm{/M}]$, $[\mathrm{C/M}]$, and $[\mathrm{N/M}]$ abundances. Second, the pipeline derives elemental abundances by tuning the fit around spectral windows of individual elements. The stellar parameter and abundance estimates are available in the \texttt{FPARAM} and \texttt{FELEM} arrays respectively. ASPCAP also calibrates these raw estimates to provide the final estimates. It is important to note that abundances are internally calibrated to ensure that clusters are chemically homogeneous.

\begin{figure}[t]
  \centering
  \includegraphics[width=\linewidth]{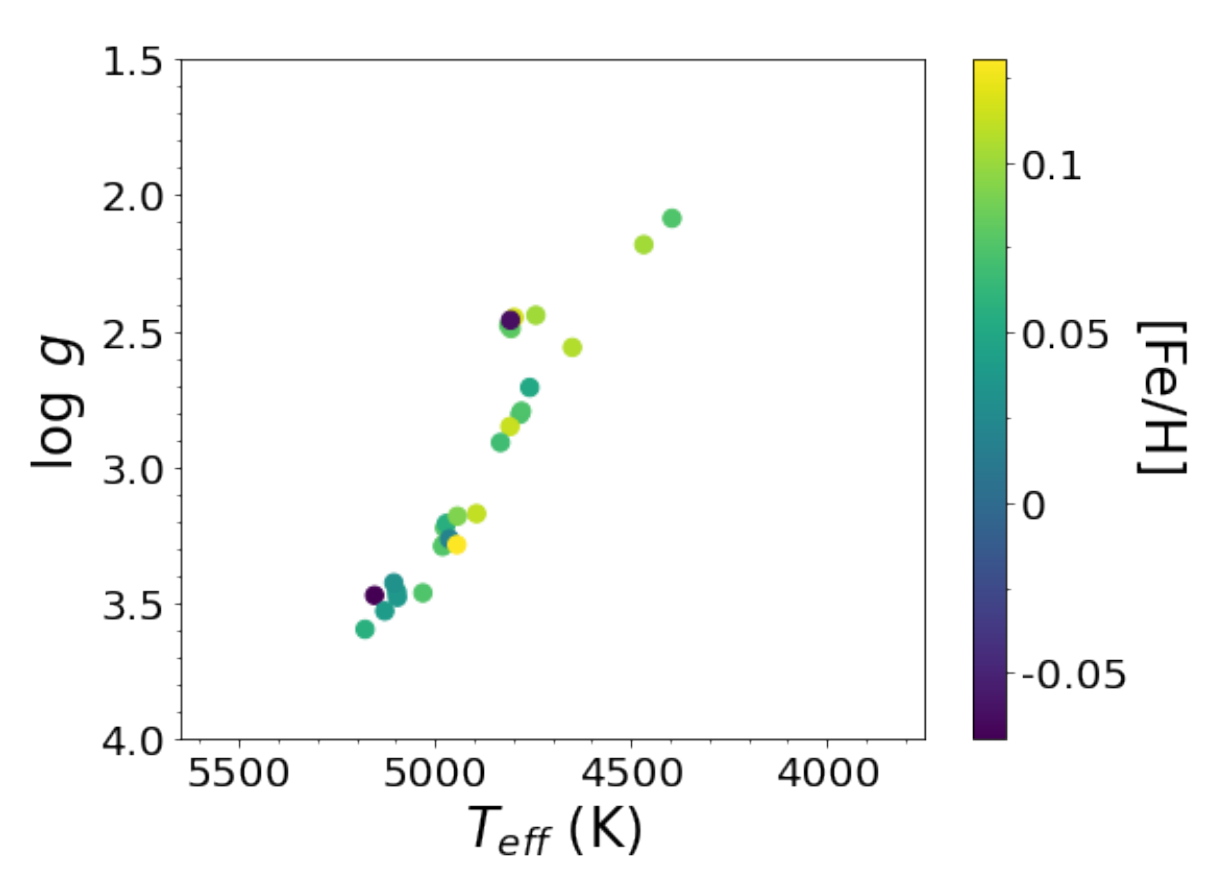}
  \caption{Spectroscopic H-R diagram ($\log g$ vs. $T_{\mathrm{eff}}$) of the 28 giant members in the OCCAM sample of open cluster M67 (Refer to Section \ref{subsec:occam} for details). Each point is color-coded based on its metallicity $[$Fe/H$]$. The $T_{\mathrm{eff}}$, $\log g$, and $[$Fe/H$]$ estimates are ASPCAP-calibrated fits. We use these members in Section \ref{sec:M67} for investigating the chemical homogeneity in M67.}
  \label{fig:M67}
\end{figure}

\subsection{Stellar sample}\label{subsec:sample}
The initial sample for this work includes all APOGEE giant members ($\log g < 4$) having metallicity within a range centered around solar metallicity ($-0.15 < $ $[$Fe/H$]$ $ < 0.15$) and high signal-to-noise ratio ($\mathrm{S/N} > 200$). The stars in this sample are mainly red giant or red clump stars and include some subgiants. We remove a few stars from this sample based on two reasons: if their ASPCAP stellar parameter estimates do not have physical values or their percentage of masked wavelengths is greater than 50 \%. The resulting sample does not include any star that has its APOGEE flag \texttt{STAR\_BAD} set, indicating that no ``bad" stars are present. Figure \ref{fig:sample} shows the spectroscopic Hertzsprung-Russell diagram of this sample.

APOGEE targets are predominantly red giant stars because their high luminosity lets us observe them at high signal-to-noise ratio over large distances \citep{zasowski_2013, zasowski_2017}. In addition, they belong to stellar populations spanning a wide range of ages and metallicities, which makes them a good tracer population for studying the history of the Galaxy. In the context of stellar chemical space, the use of red giants has two main advantages: minimal rotational line-broadening effects due to slow rotation and surface abundances reflective of the stars' initial abundances due to minimized diffusion \citep{dotter_2017}. Therefore, we mainly use APOGEE giants in our analysis.

We set the $[$Fe/H$]$ cut based on the metallicity range of open cluster M67, which we use as an application of our method in Section \ref{sec:M67}. This cut ensures that we include several giants with metallicities close to solar that mimic those in M67.

We use the Python package \texttt{apogee}\footnote{\url{https://github.com/jobovy/apogee}} \citep{bovy_2016} to download the APOGEE data and perform cuts as described above. The sample includes a total of $18,933$ stars and we use the ``global" weighted version of their combined \texttt{apStar} spectra. We continuum-normalize the combined spectra using the \cite{ness_2015} method in \texttt{apogee}, which fits a quadratic polynomial to a set of predetermined continuum pixels.

\subsection{OCCAM Open Clusters}\label{subsec:occam}
We assess the chemical homogeneity of open cluster M67 in Section \ref{sec:M67}. For this, we obtain M67 spectra from the APOGEE/Open Cluster Chemical Abundances and Mapping (OCCAM) Data Release 14 sample \citep{donor_2018}. We use M67 stars identified as giant members (\texttt{GM}) in OCCAM using the criterion $\log g \lessapprox 3.7$. Figure \ref{fig:M67} shows the spectroscopic H-R diagram for these members using ASPCAP estimates.

\subsection{Masking and Uncertainties}\label{subsec:apogee_error}
Each pixel of an APOGEE spectrum has an associated uncertainty caused by several sources of noise. We use the reported pixel uncertainties in all our spectral analyses. The systematic errors encountered during data processing and calibration of APOGEE spectra lead to a 0.5\% uncertainty floor \citep[][]{nidever_2015}, and therefore we set any uncertainties of continuum-normalized spectra smaller than 0.005 to 0.005.

In addition, each spectrum is accompanied by an \texttt{APOGEE\_PIXMASK} bitmask that flags pixels that are heavily affected by noisy features. We mask pixels identified as erroneous using bits \texttt{BADPIX}, \texttt{CRPIX}, \texttt{SATPIX}, \texttt{UNFIXABLE}, \texttt{BADDARK}, \texttt{BADFLAT}, \texttt{BADERR}, \texttt{NOSKY}, \texttt{PERSIST\_HIGH}, \texttt{PERSIST\_MED}, \texttt{PERSIST\_LOW}, \texttt{SIG\_SKYLINE}, and \texttt{SIG\_TELLURIC} corresponding to binary digits 0, 1, 2, 3, 4, 5, 6, 7, 9, 10, 11, 12, and 13 respectively \citep{holtzman_2015}.

\cite{bovy_2016} analyzed the reported APOGEE uncertainties and found that the supposedly uncorrelated errors have significant correlations with neighboring pixels over ranges ten times larger than the reported line-spread function. These correlations are potentially due to the continuum normalization of spectra. They also deduced that the errors are often underestimated by 10\%-20\% and can scale up to 100\% for several wavelength regions, especially those dominated by telluric absorption lines. Therefore, care needs to be taken while using the reported errors in analyses.

\begin{figure*}[t]
    \centering
    \includegraphics[width=\linewidth]{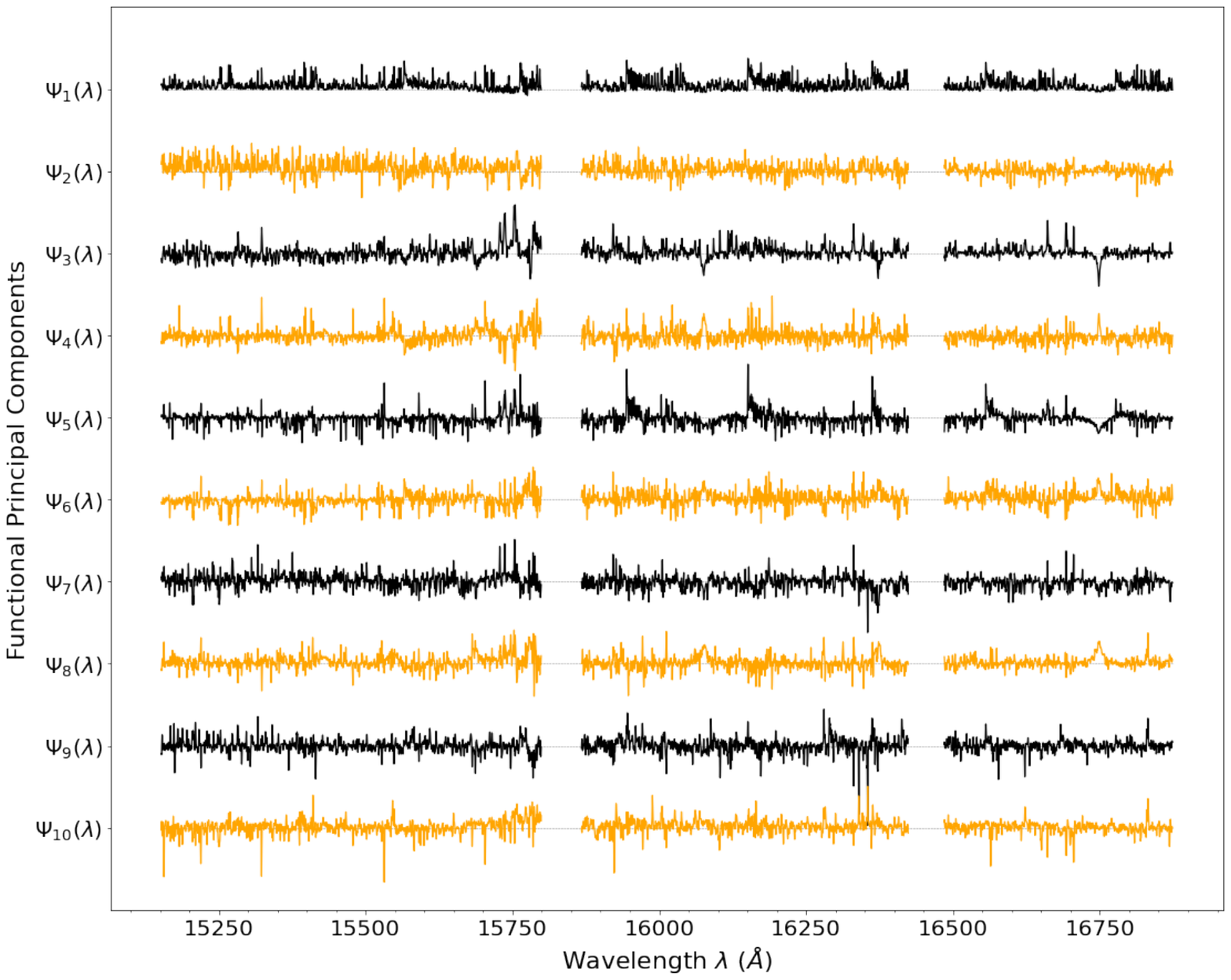}
    \caption{Functional principal components (or eigenfunctions $\Psi(\lambda)$) of the APOGEE stellar sample described in Section \ref{subsec:sample}. The panel shows the first 10 FPCs. We compute these using the method described in Section \ref{subsec:FPCA_spectra}. It is clear that these FPCs are dominated by narrow, absorption-like features, without any obvious systematics. A detailed interpretation of these FPCs is provided in Section \ref{subsec:interp_fpc}.}
    \label{fig:fpca_comp}
\end{figure*}

\section{Dimensionality of spectra}\label{sec:dim_spec}
High-resolution spectroscopic observations carry key information about stellar photospheres: stellar parameters and elemental abundances. To simultaneously infer $n = 10 - 40$ parameters from the spectra, we need to evaluate numerous photospheric models to ensure the parameter space is fully explored. Traditionally, parameter space is explored using a grid of models. Such approaches scale exponentially as $O(2^n)$. With such a large number of models to evaluate, computational speed is of the essence. One approach to improve efficiency is to assume a one-dimensional, plane-parallel, static star with local thermodynamic equilibrium (LTE). While this assumption is acceptable from a theoretical standpoint (and other considerations), it is ultimately an approximation that often fails in practice.

Newer approaches \citep[e.g.,][]{rix_2016, ting_2019} result in increased efficiency of model evaluation. However, any direct fit to the spectra must deal with the complex systematics due to, e.g., continuum-normalization errors and persistence in the detector. As spectra are very high-dimensional (often having thousands of pixels per spectrum), it is hard to separate the part of the data that has useful information from the less useful parts, especially in the light of systematics. Therefore, disentangling the component of spectral data space intrinsic to the stars from that due to systematics is an important ingredient in modern analyses, which has not yet received much attention. In this paper, we focus on this theme and extract the intrinsic spectral structure that embeds stellar parametric and abundance information.

\begin{figure*}[ht]
    \centering
    \includegraphics[width=\linewidth]{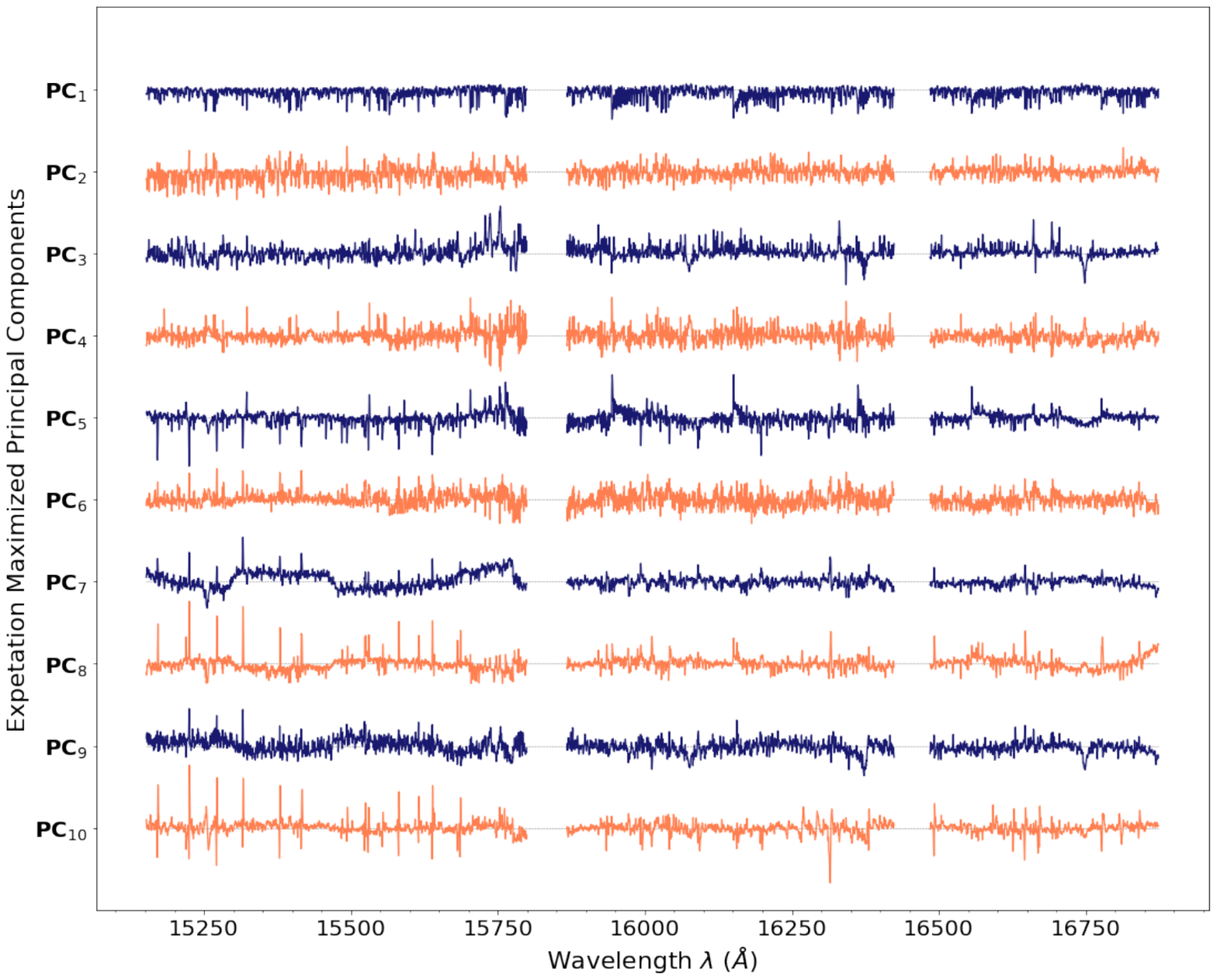}
    \caption{The first 10 expectation-maximized principal components of the APOGEE stellar sample described in Section \ref{subsec:sample}. The panel shows the first 10 EM PCs. Narrow, absorption-like features dominate the first four EM PCs. However, large-scale structure and broad, localized features become evident starting $\mathbf{PC}_5$, which we can attribute to systematics such as persistence in the detector, incorrect continuum normalization, residual sky emission and/or telluric absorption, etc. In contrast, the FPCs in Figure \ref{fig:fpca_comp} show no signs of systematics (refer to Section \ref{subsec:interp_fpc} for more details).}
    \label{fig:empca_comp}
\end{figure*}

Our APOGEE sample includes $18,933$ spectra (refer to Section \ref{subsec:sample}) and each spectrum has a wavelength grid of $7214$ pixels. The multi-element abundance variations in spectra are limited to certain wavelength regions and are heavily correlated; moreover, stellar parameters have overall effects on spectra that have correlations with some of the abundances. Thus, the intrinsic dimensionality of spectra is expected to be much smaller than $7214$. Some studies attempt to characterize the intrinsic chemical structure of spectra using inferred abundances, e.g., \cite{ting_2012pca} and \cite{andrews_2017} apply PCA to reduce the dimensionality of a set of multi-element abundances. Issues associated with the inference of abundances, however, have caused a shift toward the direct usage of spectra. For example, \cite{price_jones_2018} apply EMPCA on a subset of APOGEE spectra and find that only ${\approx}10$ orthogonal dimensions are needed to maximally explain chemical information. \cite{ting_2021} demonstrate similar results by showing that at least seven elemental abundances are required to remove cross-element correlations in spectra.

While these studies somewhat agree on the number of dimensions, the actual dimensions are up for debate. The reason for this argument is that noise and especially systematics in spectroscopic data make dimensionality reduction nontrivial. PCA is one of the most commonly used methods for dimensionality reduction; it transforms data to orthogonal dimensions that are \textit{independent} and removes collinearity in data to avoid issues in regression problems \citep{jolliffe_1986}. However, as described in Section \ref{sec:case}, PCA is unable to distinguish between noise and intrinsic variance when the sample size is small and is particularly susceptible to systematic biases. EMPCA too faces similar issues since it requires a near-perfect noise model of the data to deal with systematics. Therefore, we use FPCA to reveal the underlying structure of noisy spectra.

\begin{figure}[ht]
    \centering
    \includegraphics[width=\linewidth]{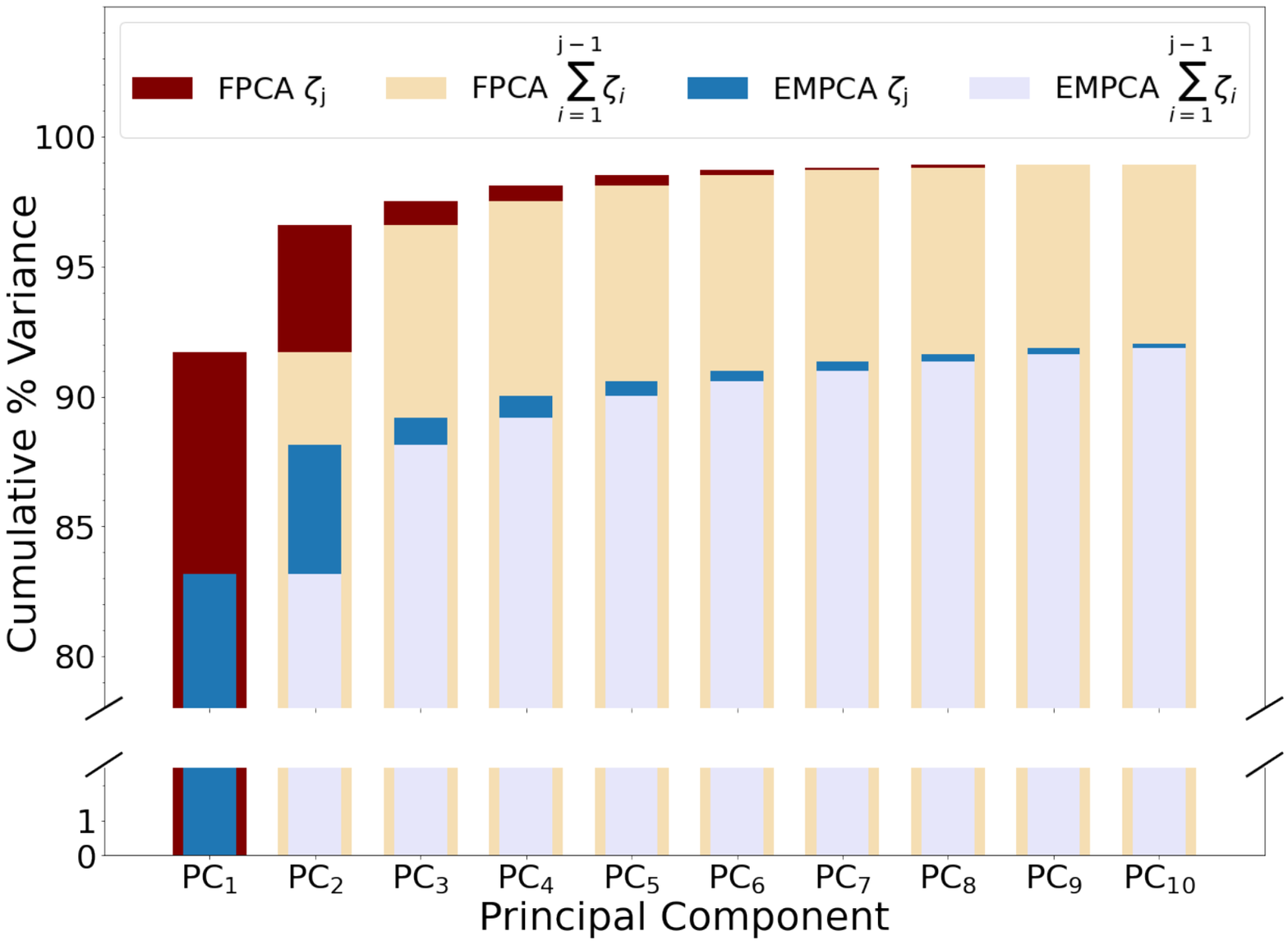}
    \caption{Cumulative percentage of total variance explained by the first 10 FPCs and EM PCs shown in Figures  \ref{fig:fpca_comp} and  \ref{fig:empca_comp} respectively. These are computed using eigenvalues as $100 \times \zeta_j, \forall j \in [1, 10]$. The maroon bars indicate the percentage explained by the corresponding FPC, $\Psi_j(\lambda)$, on the x-axis, whereas the wheat bars indicate the cumulative percentage explained by all previous FPCs, $\Psi_1(\lambda)$ to $\Psi_{j-1}(\lambda)$. The blue and lavender bars explain the same information as the maroon and wheat bars respectively, but for the EM PCs.}
    \label{fig:eigvals}
\end{figure}

\subsection{Spectral FPCA dimensions}\label{subsec:FPCA_spectra}
We follow the same procedure as in the case study (Section~\ref{subsec:case_FPCA}) to compute the FPCs of the APOGEE spectral sample described in Section~\ref{subsec:sample}. The main difference is that now $N = 18,933$ and $M = 7214$. We choose the number of basis functions $K=50$ for this sample using the stepwise variable selection approach described in Section \ref{subsec:chooseK}. To generate the basis of 50 model spectra $\phi = \{\phi_1, \dotsc, \phi_{50}\}$, we randomly select 50 spectra from the APOGEE spectral sample (refer to Section \ref{subsec:sample}) and plug their ASPCAP-calibrated stellar parameter and abundance estimates into Equation \ref{eq:basis_PSM}. We then regress our sample onto the basis functions using the weighted least-squares criterion (Equation~\eqref{eq:SSE_W_vec}). We mask the flux values before regression using the \texttt{APOGEE\_PIXMASK} bitmask described in Section \ref{subsec:apogee_error}; the mask removes pixel-level instrumental noise, persistence regions, sky lines, and telluric lines in the spectra.

\begin{figure*}[t]
    \centering
    \includegraphics[width=\linewidth]{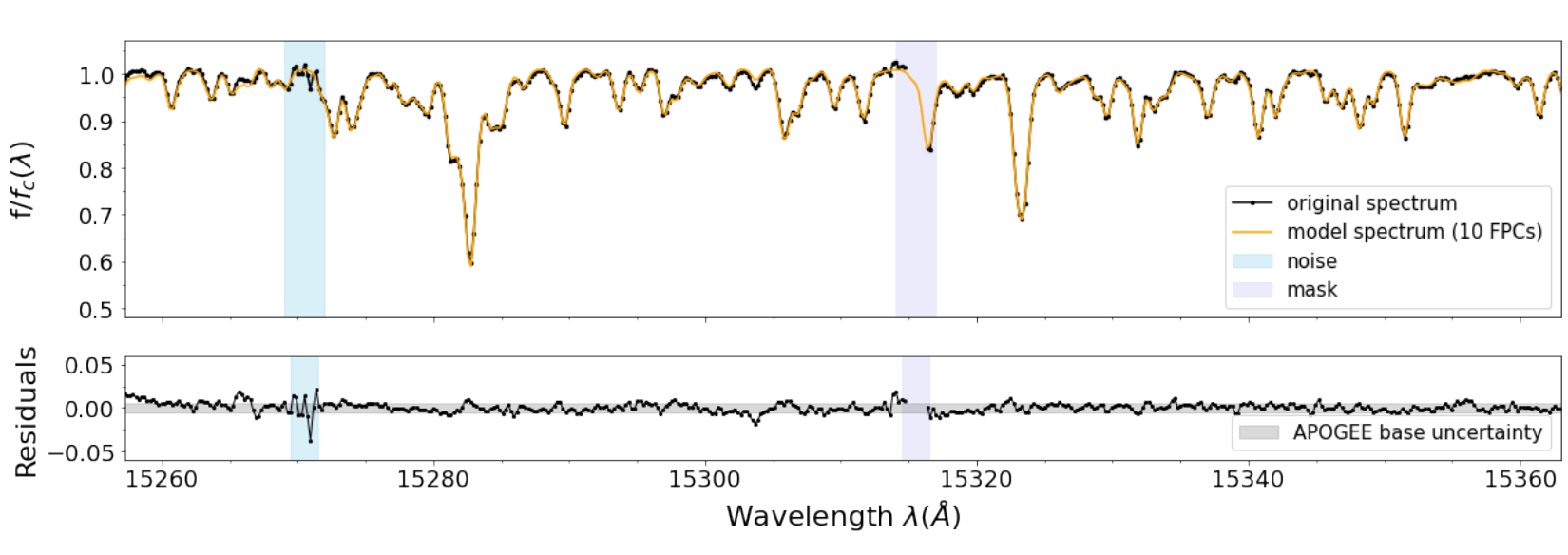}
    \caption{Modeling an M67 giant spectrum using the first 10 FPCs. The top panel shows the original spectrum and an overplotted reconstructed model. The bottom panel shows the residuals after subtracting the model from the original spectrum along with a horizontal gray region that marks the APOGEE base systematic uncertainty (0.5 \%). The two horizontal regions in both the panels highlight what FPCA does best -- remove noise and generate missing (masked) data.}
    \label{fig:fpca_model}
\end{figure*}

Given the functional representations of spectra, we compute the FPCs $\Psi(\lambda)$. Figure \ref{fig:fpca_comp} shows the first $10$ of these FPCs and Figure \ref{fig:eigvals} displays their percentage of explained variance in a cumulative fashion.

\subsection{Interpretation of Principal Components}\label{subsec:interp_fpc}
To evaluate the advantages of FPCA over classical PCA and EMPCA, we compare the FPCs shown in Figure \ref{fig:fpca_comp} with the PCs obtained using EMPCA on the same stellar spectral sample. EMPCA incorporates weights to handle noisy as well as missing data and computes PCs using the EM algorithm. In our case, we downweight spectral data points based on the APOGEE uncertainties and the \texttt{APOGEE\_PIXMASK}, for which we use the \texttt{empca}\footnote{\url{https://github.com/sbailey/empca}} implementation from \cite{bailey_2016}. We refer the reader to this paper, \cite{price_jones_2018}, and references therein for more information on EMPCA and how it applies to APOGEE spectra. Figure \ref{fig:empca_comp} shows the PCs obtained using EMPCA, which we refer to as EM PCs.

Figure \ref{fig:eigvals} compares the percentage variance explained by the FPCs with that explained by the EM PCs. The cumulative percentage explained by the first 10 FPCs is ${\approx}99\%$, whereas that explained by the first 10 EM PCs is ${\approx}91\%$. As discussed in Section \ref{subsec:case_FPCA}, the eigenvalues corresponding to FPCs are usually higher than those corresponding to classical PCs or EM PCs because the former do not quantify variance due to noise. By explaining only intrinsic variance in the data, FPCs capture physically expected patterns with fewer components.

A visual inspection of Figure \ref{fig:fpca_comp} informs us that narrow features dominate the FPCs and they do not exhibit any large-scale trends. These narrow features mainly represent absorption lines in spectra that occur at specific wavelengths, and their magnitudes indicate which lines are correlated or anticorrelated. Essentially, the FPCs capture information in spectra corresponding to stellar parameter and chemical abundance variations. In contrast, narrow absorption-like features dominate only the first few EM PCs in Figure \ref{fig:empca_comp} and large-scale structure becomes evident starting from $\mathbf{PC}_5$. Specifically, $\mathbf{PC}_7$, $\mathbf{PC}_8$, and $\mathbf{PC}_9$ have significant large-scale trends in the blue detector (``chip") of APOGEE. This detector ranges from 15140 to 15810 $\AA$, and about one-third of this range is severely affected by the instrumental effect \textit{superpersistence}. Persistence or its extreme case, superpersistence, is not uncommon in APOGEE spectra, and the \texttt{APOGEE\_PIXMASK} masks regions affected by it. However, even after masking the persistence, we are still left with obvious persistence issues. These issues affect EMPCA because it cannot distinguish between signal variance and variance due to noise if the noise characterization is imperfect (refer to Section \ref{subsec:apogee_error}). We thus observe persistence in our EM PCs similar to what we see in \cite{price_jones_2018}, where they use EMPCA on a subsample of APOGEE giants after removing stellar parametric trends using polynomial models; their EM PCs incorporate only chemical features, whereas ours include stellar parametric effects as well.

In addition to persistence, continuum-normalization systematics may also contribute to the large-scale trends seen in the EM PCs. The EM PCs exhibit some localized features as well, which are potentially due to instrumental effects. For example, $\mathbf{PC}_9$ and $\mathbf{PC}_{10}$ have a broad feature at ${\sim}16350 \AA$ and $\mathbf{PC}_6$ to $\mathbf{PC}_9$ display an upward or downward trend toward the end of the red detector. There are also some strong absorption-like features in $\mathbf{PC}_8$ and $\mathbf{PC}_{10}$ that may be due to interfering sky emission and/or telluric absorption features.

While the above comparison shows that FPCA is capable of removing large-scale trends attributed to instrumental noise, there is still the possibility that some features in the FPCs do not correspond to stellar photospheric information. To this end, we take a closer look at the FPCs and interpret the variations we see. Some of the absorption features in the FPCs are easy to identify, such as the wide hydrogen Brackett lines around $15725 \AA$, $16100 \AA$, $16400 \AA$, and $16800 \AA$ \citep{shetrone_2015}. The FPCs also show narrow features around $15,273 \AA$, a known diffuse interstellar band (DIB) in APOGEE spectra \citep{zasowski_2015}. However, we find that the EM PCs have higher magnitudes at $15273 \AA$ than FPCs; this suggests that FPCA has at least partially reduced the effect of the DIB, a nonphotospheric feature.

We also analyze the FPCs quantitatively using two different approaches. First, we juxtapose the FPCs with a synthetic spectrum and confirm that the narrow features in the FPCs do correspond to absorption peaks in the spectrum; this comparison is similar to that in \cite{price_jones_2018}. Second, we validate that the FPCs filter the intrinsic variations in spectra by correlating them with classical PCs of a sample of mock APOGEE spectra. We generate this mock sample using the ASPCAP-calibrated stellar parameter and abundance estimates of our spectral sample. Table \ref{tab:correlation} displays the correlations in percentage form as $|100 r|$, where $r$ is the Pearson correlation coefficient, along with the percentage correlations between the EM PCs and the classical PCs. The table indicates that the FPCs have higher correlations with theoretically expected PCs than the EM PCs. It also highlights that the EM PCs: $\mathbf{PC}_{8}$, $\mathbf{PC}_{9}$, and $\mathbf{PC}_{10}$ have significantly low correlations with classical PCs, which validates our visual inspection that noise dominates them.

\begin{deluxetable}{lcc}
\tablecolumns{3}
\tablewidth{0pt}
\tablecaption{\label{tab:correlation} Correlation with Theoretical PCs}
    \tablehead{
    & \colhead{FPC Correlation (\%)}  & \colhead{EM PC Correlation (\%)}}
    \startdata
    $\mathbf{PC}_{1}$  & 94.4  & 90.3\\
    $\mathbf{PC}_{2}$  & 95.8  & 90.3\\
    $\mathbf{PC}_{3}$  & 87.8  & 75.8\\
    $\mathbf{PC}_{4}$  & 71.9  & 54.3\\
    $\mathbf{PC}_{5}$  & 66.4  & 50.1\\
    $\mathbf{PC}_{6}$  & 73.1  & 38.7\\
    $\mathbf{PC}_{7}$  & 55.8  & 24.5\\
    $\mathbf{PC}_{8}$  & 8.6   & 0.7\\
    $\mathbf{PC}_{9}$  & 25.5  & 7.8\\
    $\mathbf{PC}_{10}$ & 11.4  & 0.6\\
    \enddata
    \tablecomments{Correlation percentages between FPCs in Figure \ref{fig:fpca_comp} and the first 10 classical PCs of mock APOGEE spectra are compared with those between EM PCs (Figure \ref{fig:empca_comp}) and classical PCs. FPCs are more correlated with theoretically expected PCs than corresponding EM PCs, especially $\mathbf{PC}_{8}$, $\mathbf{PC}_{9}$, and $\mathbf{PC}_{10}$.}
\end{deluxetable}
In Section \ref{sec:M67}, we use our FPCs to constrain the chemical homogeneity in the open cluster M67, which further solidifies our argument that FPCA provides a good basis for dimensional reduction of spectra.

\subsection{Dimensionality reduction using FPCA scores}
We obtain dimensionality-reduced scores $sc_{nj}$ of an APOGEE spectrum $\mathbf{y}_n$ by projecting it onto the FPCs $\Psi_j(\lambda)$. We project using the EM algorithm to properly downweight masked and noisy regions of $\mathbf{y}_n$. Since we use a large sample of giant spectra for computing the FPCs, we expect that any star \textit{similar} to this sample can be projected onto the FPC basis. As an illustrative example, we project the spectrum of an M67 giant member (refer to Section \ref{subsec:occam}) onto the FPCs, and optimize the number of dimensions $J$ based on the discussion in Section \ref{subsec:chooseK}. We find that ${\approx}10$ dimensions are required to explain total variance up to the level of the base systematic error in APOGEE, and ${\approx}50$ dimensions are required to include all information on the star while excluding measurement noise. This finding agrees with Figure \ref{fig:eigvals}, which shows that the first 10 FPCs explain ${\sim}99\%$ variance, with $\Psi_9$ and $\Psi_{10}$ contributing negligible variance. Therefore, we use $10$ FPCs and compute a dimensionally reduced approximation of the M67 spectrum, which is shown as the ``model spectrum" in Figure~\ref{fig:fpca_model}. This figure highlights a few features of modeling using FPCA: (1) It is capable of detecting noise in data and removing it from the procedure because it represents extrinsic variance (for example, the sky-blue region labeled ``noise" shows variations in the top panel, which are potentially due to the DIB at $15273 \AA$, and FPCA removes these variations, as is evident from the residuals in the bottom panel), and (2) it can impute masked regions of a spectrum (e.g. the lavender region labeled ``mask").

\section{Application: Chemical Homogeneity of Open Cluster M67}\label{sec:M67}
In this section, we dimensionally reduce the spectra of giant members of open cluster M67 using our FPCs (refer to Section~\ref{subsec:occam} for details of M67 members). We then compute stellar parameters and abundances of these stars from the low-dimensional FPC scores and estimate the level of chemical homogeneity in M67.

\subsection{Introduction}\label{subsec:intro_chem}
Stars form in groups in giant molecular clouds (GMCs) \citep{shu_1987, lada_2003}. Simulations indicate that these clouds are well mixed because of turbulent mixing  \citep{feng_2014}, and this implies that stars in clusters are chemically homogeneous at birth. However, core-collapse (Type II) supernovae can enrich the intracloud medium over timescales of a few million years. These enrichment timescales are comparable to those of star-formation, which likely causes variations in birth chemistry. Beyond birth chemistry, the chemical composition of stars evolves due to stellar atmospheric processes such as deep, convective mixing and atomic diffusion. 

Several studies on the chemical abundances of open and globular clusters suggest homogeneity \citep[e.g.,][]{de_silva_2006, de_silva_2007b, de_silva_2007a, reddy_2012, ting_2012chem, bovy_2016, poovelil_2020}. In addition, \cite{kos_2020} demonstrate that the Orion complex, a nearby star-forming region, is chemically homogeneous, which suggests that core-collapse supernovae from the old clusters did not pollute the young clusters. However, many studies have found contradictory results. Observations show that globular clusters have significant dispersion in their light element abundances \citep{briley_1996, carretta_2009, meszaros_2015}. Abundances of main-sequence and turnoff stars in metal-poor globular clusters show a lack of homogeneity \citep{korn_2007, lind_2008, nordlander_2012}. Some studies have found significant inhomogeneities in M67 member stars and attributed them to stellar evolution, particularly atomic diffusion \citep{onehag_2014, liu_2019, souto_2019}.

The above stated disagreements are difficult to resolve due to instrumental effects and theoretical systematic biases in spectroscopic analyses. As described in Section \ref{sec:dim_spec}, FPCA can tackle such issues with spectral data. Using the low-dimensional space of FPCs, we can accurately and efficiently infer surface abundances of stars in M67, and constrain the chemical homogeneity in the cluster.

One reason for determining whether or not open clusters like M67 are chemically homogeneous is that, if they are, it improves prospects for strong chemical tagging \citep{freeman_2002}. Dissolved clusters within the Milky Way disk can be chemically tagged to their birth GMCs if clusters have unique chemical signatures that are homogeneous. Validating chemical homogeneity will thus allow us to map the detailed chemodynamical evolution of the Milky Way beyond our current understanding \citep{price_jones_2020}.

\subsection{Likelihood-free Inference of Abundances}\label{subsec:LFI}
Likelihood-free inference (LFI) refers to the inference of parameters of a statistical model that can forward-model data but whose likelihood function is intractable. Such stochastic models (also called simulator-based models) provide flexible ways to describe complex physical processes in astronomy; however, the lack of their likelihood functions makes it difficult to infer model parameters $\boldsymbol{\theta}$ given observed data $\mathbf{d}$ since posterior
probability distribution functions (PDFs) $ p(\boldsymbol{\theta}\mid \mathbf{d})$ are estimated using Bayes' rule
\begin{equation}\label{eq:bayes}
\begin{split}
    p(\boldsymbol{\theta}\mid \mathbf{d}) &= \frac{p(\mathbf{d}\mid\boldsymbol{\theta})\, p(\boldsymbol{\theta})}{p(\mathbf{d})} \\
    posterior &\propto likelihood \times prior.
\end{split}
\end{equation} Even though the exact evaluation of a likelihood is not possible, LFI approximates posterior PDFs using simulated data.

\cite{bovy_2016} developed a method using approximate Bayesian computation (ABC) \citep{tavare_1997}, a traditional LFI technique, to estimate the initial abundance variation of 15 elements in open clusters M67, NGC 6819, and NGC 2420. Evaluating the exact likelihood of stellar spectra requires an accurate model of spectral noise that is unavailable (refer to Section \ref{subsec:apogee_error}) -- this motivates the use of ABC, which estimates posterior PDFs using simulated data that match the observed data based on an $\epsilon$ criterion. However, the drawback of ABC is that the number of simulations required dramatically increases as $\epsilon \to 0$, especially when the dimensionality of data is high. Therefore, ABC is infeasible when the cost of simulation is even moderately expensive.

Sequential neural likelihood \citep[SNL;][]{papamakarios_2019sequential, lueckmann_2019} is a new $\epsilon$-free LFI method through which one can sequentially train a flexible \textit{conditional neural density estimator} (NDE) to learn a likelihood using orders-of-magnitude fewer simulations than traditional ABC methods. NDEs learn parametric approximations of complex likelihoods, whereas the sequential procedure adaptively acquires more simulations in the region of high posterior density. SNL has successful applications in several science and engineering fields; particularly interesting applications are those in cosmology \citep{alsing_2019}.

SNL falls into the broader category of neural ($\epsilon$-free) LFI for simulation-based models along with methods such as sequential neural posterior estimation \citep[SNPE;][]{papamakarios_2016, lueckmann_2017} and sequential neural ratio estimation \citep[SNRE;][]{durkan_2020, hermans_2020}. SNPE directly approximates the posterior using NDEs, whereas SNRE uses neural-network-based classifiers to estimate density ratios that are proportional to the likelihood. Learning the likelihood using SNL instead of the posterior using SNPE is often easier, and has the advantage that the proposal (updated prior in the sequential procedure) does not introduce bias into the approximation \citep{papamakarios_2019sequential}. \cite{durkan_2020} perform experiments to show that SNL also outperforms SNRE; they highlight that SNRE is advantageous when the dimensionality of data is high since it uses a classifier that takes the data as input rather than parameterizing a high-dimensional distribution.

The amount of data (simulations) required to perform density estimation grows exponentially as dimensionality increases, and this is generally referred to as the \textit{curse of dimensionality} in machine learning. \cite{papamakarios_2019thesis} illustrates this problem faced by density estimation using a simple example. Due to the overall advantages of SNL over SNRE, we use it here along with FPCA dimensionality reduction.

The goal is to infer the stellar parameters and abundances $\boldsymbol{\theta}$ of each M67 giant star in our sample by estimating its posterior PDF $p(\boldsymbol{\theta}\mid \mathbf{d}_{ob})$, where $\mathbf{d}_{ob}$ is the observed APOGEE stellar spectrum (Refer to Section~\ref{subsec:occam} for details of the sample). We use the inference procedure in \cite{papamakarios_2019sequential}, with some modifications introduced in \cite{alsing_2019}. Figure \ref{fig:flowchart} illustrates this procedure, which we separately perform for each star. The inputs to this procedure are:

\begin{figure}[t]
  \includegraphics[trim=70 5 90 2, clip, width=\linewidth]{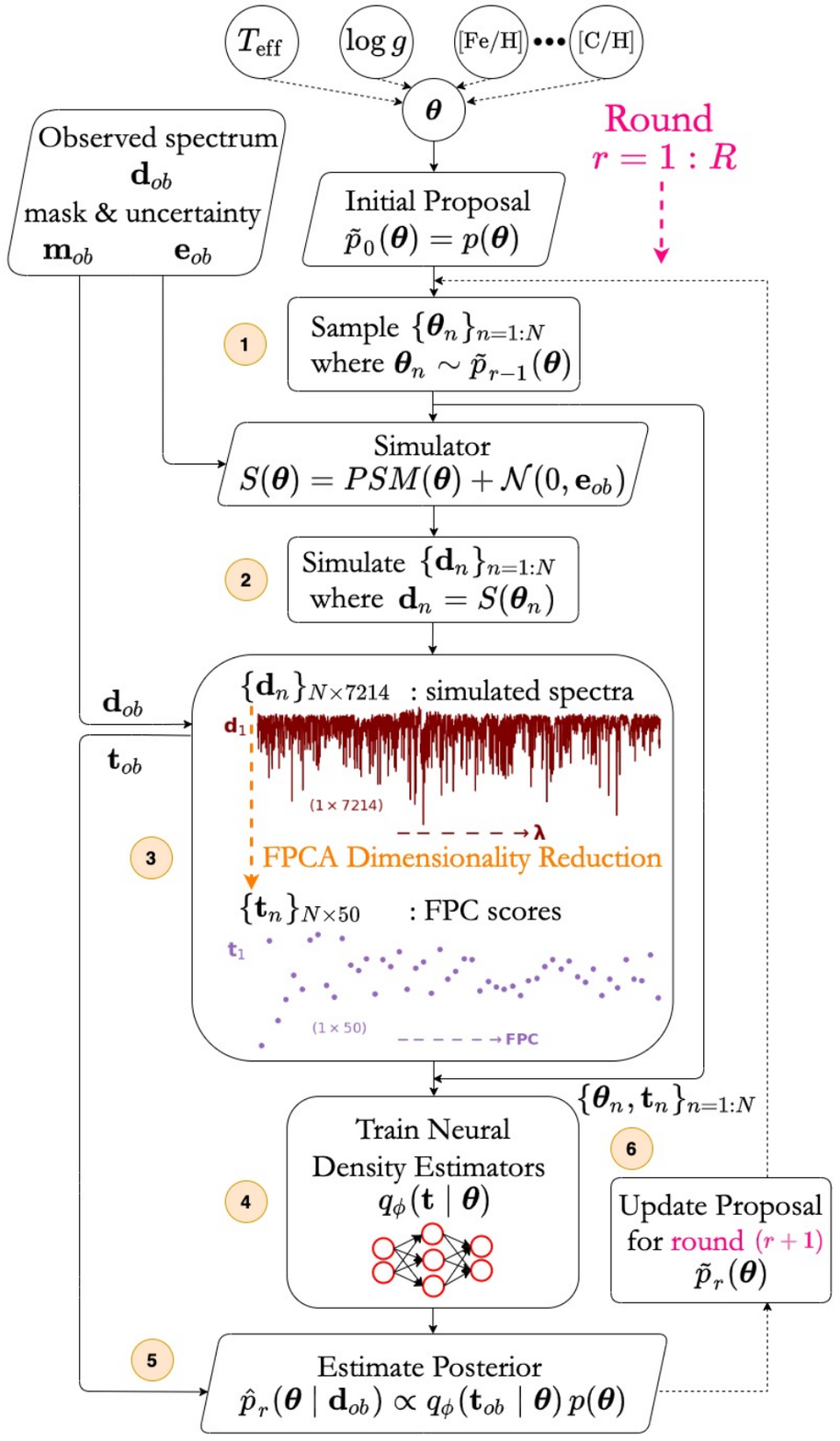}
  \caption{Graphical representation of spectroscopic inference using sequential neural likelihood for each M67 giant star used in this analysis (refer to Section \ref{subsec:LFI}).}
  \label{fig:flowchart}
\end{figure}

\begin{enumerate}
    \item \textit{Simulator}: $S(\boldsymbol{\theta}) = PSM(\boldsymbol{\theta})+ \mathcal{N}(0, \mathbf{e}_{ob})$\\ This generates an APOGEE spectrum given model parameters $\boldsymbol{\theta} = \{T_\mathrm{eff},\:log\:g\} \cup \{\:[\mathrm{X/H}] \mid \mathrm{X} =\{\mathrm{C,\:N,\:O,\:Na,\:Mg,\:Al,\:Si,\:S,\:K,\:Ca,\:Ti,\:V,\:Mn,}\\ \mathrm{Ni,\:Fe}\}\}$. We simulate using the PSM spectroscopic model discussed in Section~\ref{subsec:case_FPCA} \citep{rix_2016}. To mimic noisy data, we add Gaussian noise using the APOGEE reported uncertainties $\mathbf{e}_{ob}$ of the observed spectrum $\mathbf{d}_{ob}$.
    \item \textit{Initial proposal}: $\widetilde p_o(\boldsymbol{\theta}) = p(\boldsymbol{\theta})$\\ This specifies the initial proposal PDFs on model parameters. We use the uniform priors in Table \ref{tab:prior} for the initialization.
    \item \textit{Observed spectrum}: $\mathbf{d}_{ob}$,\\
    \textit{and corresponding mask and uncertainties}: $\mathbf{m}_{ob}, \; \mathbf{e}_{ob}$\\
    These provide the observed APOGEE spectrum of the M67 star whose parameters we want to infer. We mask this spectrum based on the criteria described in Section \ref{subsec:apogee_error}, and use uncertainties from the APOGEE catalog.
\end{enumerate}

\begin{deluxetable}{ll}
\centering
\tablecolumns{2}
\tablecaption{\label{tab:prior}Prior PDFs $p(\boldsymbol{\theta})$ for SNL}
\tablehead{
    \colhead{Model Parameter} \hspace{3cm} & \colhead{Prior}}
    \startdata
        $T_\mathrm{eff}$ & $\mathcal{U}(4000, 5500)$ \\
        $\log g$ & $\mathcal{U}(0, 4)$ \\
        $[\mathrm{Fe/H}], \, [\mathrm{N/H}]$ & $\mathcal{U}(-1, 1)$\\
        All other $[\mathrm{X/H}]$ & $\mathcal{U}(-0.5, 0.5)$\\
    \enddata
    \tablecomments{This table gives priors $p(\boldsymbol{\theta})$ on the parameters $\boldsymbol{\theta} = \{ T_\mathrm{eff},\,\log g,\,\left[\mathrm{X/H}\right]\}$ where $\mathrm{X} = \{ \mathrm{C},\,\mathrm{N},\,\mathrm{O},\,\mathrm{Na},\mathrm{Mg},$\\$\;\;\mathrm{Al},\,\mathrm{Si},\,\mathrm{S},\,\mathrm{K},\,\mathrm{Ca},\,\mathrm{Ti},\,\mathrm{V},\,\mathrm{Mn},\,\mathrm{Fe},\,\mathrm{Ni}\}$. These are used in the SNL procedure described in Section \ref{subsec:LFI}.}
\end{deluxetable}

Given the above inputs, the inference procedure is run over multiple rounds to \textit{sequentially} approximate the likelihood function of the observed APOGEE spectrum. In each round, indexed by $r=1, \dotsc, R$, we run the following steps (corresponding to the numbered boxes in Figure \ref{fig:flowchart}):
\begin{enumerate}
    \item \textit{Sample $\{\boldsymbol{\theta}_n\}_{n=1:N}$ where $\boldsymbol{\theta}_n\ \sim \widetilde p_{r-1}(\boldsymbol{\theta})$}.\\
    We sample a batch of $N$ model parameters $\{\boldsymbol{\theta}_n\}_{n=1:N}$ from the proposal $\widetilde p_{r-1}(\boldsymbol{\theta})$.
    
    \item \textit{Simulate $\{\mathbf{d}_n\}_{n=1:N}$ where $\mathbf{d}_n = S(\boldsymbol{\theta}_n)$}.\\
    We then simulate spectra $\{\mathbf{d}_n\}_{n=1:N}$ corresponding to the sampled parameters $\{\boldsymbol{\theta}_n\}_{n=1:N}$ using the \textit{simulator} $S(\boldsymbol{\theta})$.
    
    \item  \textit{$\{\mathbf{d}_n\}_{N \times 7214} \rightarrow \{\mathbf{t}_n\}_{N \times 50}$ using FPCA}.\\
    In order to perform fast inference using NDEs, we reduce the dimensionality of the simulated spectra $\{\mathbf{d}_n\}_{N \times 7214}$ from $7214$ to $50$ by projecting them onto the FPCs computed in Section \ref{subsec:FPCA_spectra}. We use the EM algorithm to incorporate the mask $\mathbf{m}_{ob}$ and uncertainties $\mathbf{e}_{ob}$ of the observed spectrum in the projection. Using the same method, we dimensionally reduce the observed spectrum $\mathbf{d}_{ob} \rightarrow \mathbf{t}_{ob}$.
    
    \item \textit{Train NDEs $q_{\phi}(\mathbf{t} \mid \boldsymbol{\theta})$ to learn the likelihood}.\\
    Inspired by the ensembles in \cite{alsing_2019}, we use a masked autoencoder for distribution estimation \citep[MADE,][]{germain_2015} along with multiple mixture density networks \citep[MDNs,][]{bishop_1994} to estimate an approximate likelihood; this incorporates multiple network architectures for robust and accurate density estimation \citep{lakshminarayanan_2017}. We (re-)train this ensemble on a total of $rN$ $\{\boldsymbol{\theta}_n, \mathbf{t}_n\}$ samples generated in rounds $1, \dotsc, r$. Section \ref{subsec:LFIimplement} provides details of the NDEs and their training.
    
    \item \textit{Estimate posterior $\hat p_r(\boldsymbol{\theta} \mid \mathbf{d}_{ob}) \propto q_{\phi}(\mathbf{t}_{ob} \mid \boldsymbol{\theta}) p(\boldsymbol{\theta})$}.\\
    We estimate the desired posterior PDF $\hat p_r(\boldsymbol{\theta} \mid \mathbf{d}_{ob})$ using the learned likelihood at the dimensionally reduced spectrum $q_{\phi}(\mathbf{t}_{ob} \mid \boldsymbol{\theta})$.
    
    \item \textit{Update proposal for round (r+1) $\widetilde p_r(\boldsymbol{\theta})$}.\\
    The proposal controls what region of parameter space is sampled for learning the likelihood $p(\mathbf{d}\mid\boldsymbol{\theta})$. Since our aim is to estimate the posterior for a \textit{specific} observation, $p(\boldsymbol{\theta}\mid \mathbf{d}_{ob})$, a proposal that is concentrated in regions of high posterior density will reduce the number of simulations required for the inference. Therefore, the proposal for round $r+1$ $\widetilde p_{r}(\boldsymbol{\theta})$ is set using the approximate posterior of the previous round as $\sqrt{\hat p_{r}(\boldsymbol{\theta} \mid \mathbf{d}_{ob}) p(\boldsymbol{\theta})}$. This adaptive learning strategy is from \cite{alsing_2019}; it differs slightly from \cite{papamakarios_2019sequential} who set the proposal as $\widetilde p_{r}(\boldsymbol{\theta}) = \hat p_{r}(\boldsymbol{\theta} \mid \mathbf{d}_{ob})$.
\end{enumerate}

\begin{figure*}[t]
  \centering
  \includegraphics[width=\textwidth]{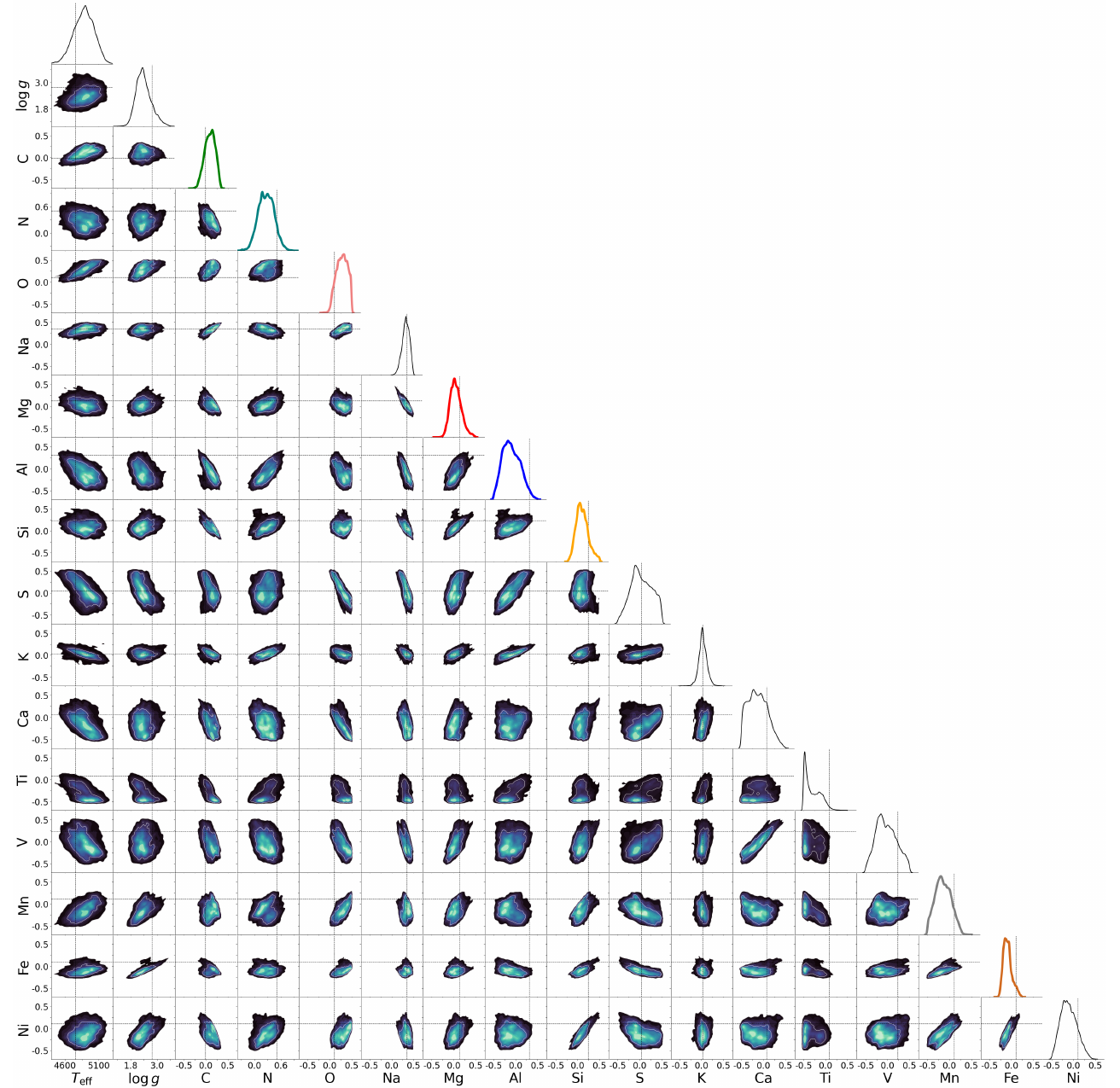}
  \caption{The posterior PDF of stellar parameters ($T_\mathrm{eff}, \log g$) and 15 elemental abundances ($\mathrm{X/H}$) of the M67 giant star, 2M08511710+1148160, given its APOGEE spectrum, is displayed as a \textit{triangle or corner plot}. The inner white contour represents the 68\% credible interval, whereas the outermost boundary is at 95\%. Dashed black lines represent the APOGEE estimated stellar parameters and chemical abundances provided in the \texttt{FPARAM} and \texttt{FELEM} data fields respectively (refer to Section \ref{sec:data} for details of APOGEE estimates). Marginal distributions are shown in the diagonal; some of them are highlighted using different colors to draw attention toward the abundances we are interested in.}
  \label{fig:M67postsingle}
\end{figure*}

\subsection{Implementation Details}\label{subsec:LFIimplement}
To perform the SNL inference, we use \texttt{pydelfi}\footnote{\url{https://github.com/justinalsing/pydelfi}} \citep{alsing_2019}, a Python package based on methods developed in \cite{papamakarios_2019sequential}, \cite{lueckmann_2019} and \cite{alsing_2018compression}. Using this package, we create an ensemble of NDEs that learns the likelihood function of the M67 spectra. The ensemble consists of up to five NDEs: a MADE with two hidden layers of 10 units each along with a combination of one, two, three, and four component MDNs, each with two hidden layers made up of 15 units; the activation functions for all the hidden layers are \texttt{tanh}. We tune the combination of MDNs for each spectrum to ensure that the network architecture is robust despite varying signal-to-noise ratio and masking of spectra.

We train the NDEs using the \texttt{Adam} \citep{kingma_2015} stochastic gradient-based optimizer, with a minibatch size of $100$ and $0.001$ learning rate. We use 600 simulations in the first round of SNL followed by 300 simulations in each subsequent round. To prevent overfitting, we employ early-stopping, i.e., we compute the validation log-likelihood over 10\% of the simulations used in each SNL round, and stop training if it does not improve after 10 epochs. We run the SNL procedure individually for the 28 giants in our sample and find that each of them converges within $O(10^3)$ simulations. Training and validation loss help us track convergence.

\begin{figure*}[ht]
    \includegraphics[width=\linewidth]{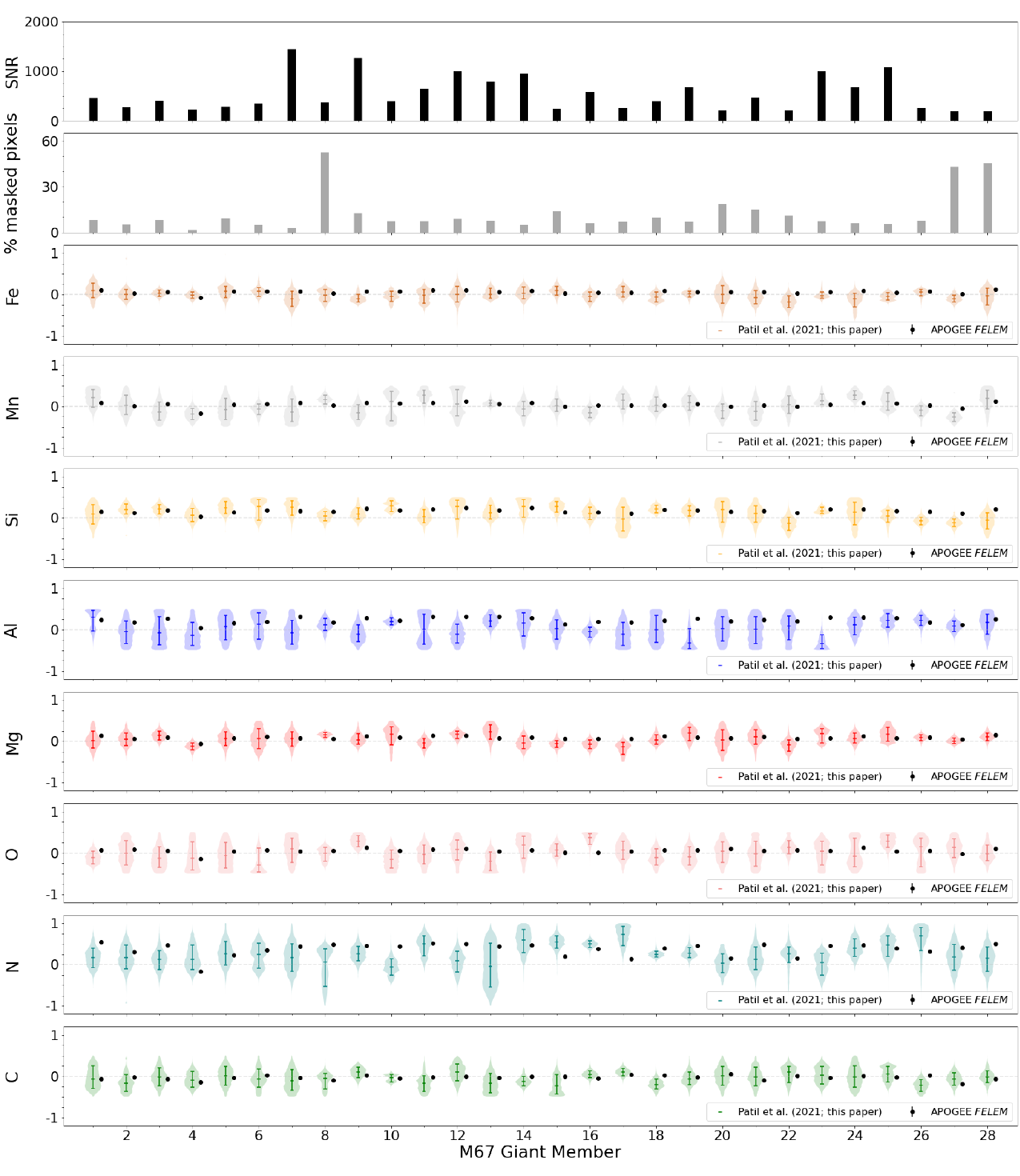}
    \caption{Inferred marginal posterior PDFs of chemical abundances, $[$Fe/H$]$, $[$Mn/H$]$, $[$Si/H$]$, $[$Al/H$]$, $[$Mg/H$]$, $[$O/H$]$, $[$N/H$]$ and $[$C/H$]$, given APOGEE spectra of M67 giant stars (bottom eight panels). We display the posterior distributions as violin plots with markers to indicate the median and the 68 \% credible interval, and they are color-coded according to Figure \ref{fig:M67postsingle}. The distributions are compared with the APOGEE \texttt{FELEM} fits (refer to Section \ref{sec:data} for details of APOGEE fits). The top two panels show the signal-to-noise ratio and the percentage of masked pixels of the stars whose abundances have been inferred to give a sense of which stars have noisy or heavily masked spectra.}
    \label{fig:abun_post}
\end{figure*}

\subsection{Results}\label{subsec:LFIresults}
Figure \ref{fig:M67postsingle} shows the inferred posterior PDFs of stellar parameters $T_\mathrm{eff}$ and $\log g$, and 15 elemental abundances of the M67 giant, 2M08511710+1148160. The diagonal panels in the figure display marginal distributions; we highlight those of interest, $[$C/H$]$, $[$N/H$]$, $[$O/H$]$, $[$Mg/H$]$, $[$Al/H$]$, $[$Si/H$]$, $[$Mn/H$]$, and $[$Fe/H$]$, in green, teal, pink, red, blue, yellow, gray, and rust colors respectively.

Table \ref{tab:M67abundances} in the Appendix lists the 15 abundances we infer for our sample of 28 giants in M67. Figure \ref{fig:abun_post} shows the marginalized posterior PDFs of the elements of interest in these stars along with their APOGEE \texttt{FELEM} fits (refer to section \ref{sec:data} for details of APOGEE fits). The posteriors of each abundance are color-coded according to Figure \ref{fig:M67postsingle}. The top two panels display the signal-to-noise ratio and percentage of masked pixels (wavelengths) of the individual stars to indicate data quality. It is evident in the plot that some level of homogeneity is present across the inferred stellar abundances.

To properly constrain the chemical homogeneity of M67 using our posterior PDFs for the abundances of the 28 giants in M67, we use a hierarchical Bayesian model described in \cite{hogg_2010}. We combine individual stellar abundance estimates, specifically the abundance posteriors in our case, and infer the \textit{true} distribution of abundances within M67 as a whole. Given samples of posterior PDFs of a stellar abundance $[$X/H$]$ for the 28 giants, we construct a parametric model $f_{\boldsymbol\omega}\left( [\mathrm{X/H}] \right)$ for the true $p\left([\mathrm{X/H}]\right)$ in M67. Essentially, we forward-model the posterior samplings and fit for the distribution parameters $\boldsymbol\omega$ through heteroskedastic deconvolution of nuisance parameters -- this means that the convolution of $f_{\boldsymbol\omega}\left([\mathrm{X/H}]\right)$ with nuisance parameter distributions describes the $[$X/H$]$ samplings. The mathematical formulation of such hierarchical modeling is described below.

The marginal likelihood $\mathcal{L_{\boldsymbol{\omega}}}$ for parameters $\boldsymbol{\omega}$ of $f_{\boldsymbol\omega}([\mathrm{X/H}])$ is given by

\begin{equation}
    \mathcal{L_{\boldsymbol{\omega}}} \equiv p(\{\mathbf{y}_n\}_{n=1}^N \mid \boldsymbol{\omega})
\end{equation} where $\{\mathbf{y}_n\}_{n=1}^N$ are the $N=28$ spectra in M67. The sampling approximation to the likelihood is
\begin{equation}
    \mathcal{L_{\boldsymbol\omega}} \approx \prod_{n=1}^{N} \frac{1}{K} \sum_{k=1}^{K} \frac{f_{\boldsymbol\omega}([\mathrm{X/H}]_{nk})}{p_o([\mathrm{X/H}]_{nk})}
\end{equation} where $[\mathrm{X/H}]_{nk}$ are the $K=100,000$ posterior samples of each $[\mathrm{X/H}]_{n}$, $f_{\boldsymbol\omega}$ is the M67 abundance distribution we aim to infer, and $p_o$ is the uniform prior on $[$X/H$]$, which we describe in Table \ref{tab:prior}.

We assume the abundance distributions are generalized Student’s $t$-distributions, i.e., $f_{\boldsymbol\omega} = f_{\nu, \hat\mu, \hat\sigma}$, where $\nu$ is the number of degrees of freedom or normality parameter, $\hat\mu$ is the location parameter, and $\hat\sigma$ is the scale parameter. We estimate $\boldsymbol{\omega} = \nu, \hat\mu, \hat\sigma$ by sampling the posterior $\mathcal{L_{\boldsymbol\omega}} \, p(\boldsymbol{\omega})$, where $p(\boldsymbol{\omega})$ is the prior, which we assume to be uniform:
\begin{equation}
    \begin{split}
    p(\nu) &= \mathcal{U}(1, 10)\\
    p(\hat\mu) &= \mathcal{U}(-0.5, 0.5)\\
    p(\hat\sigma) &= \mathcal{U}(0, \infty).
    \end{split}
\end{equation} We carry out sampling using the Markov Chain Monte Carlo (MCMC) implementation in the \texttt{emcee}\footnote{\url{https://github.com/dfm/emcee}} Python package \citep{foreman_mackey_2013}.

\begin{deluxetable}{cC}
    \tablecolumns{2}
    \tablewidth{\linewidth}
    \tablecaption{\label{tab:abundancescatter} M67 Abundance Scatter}
    \tablehead{
        \colhead{Abundance X} & \colhead{\hspace{0.75cm} $68^\mathrm{th}$ Percentile of $f_{\nu, \hat\mu, \hat\sigma}([\mathrm{X/H}])$ (dex) \hspace{0.75cm}}\\
        & \colhead{\hspace{0.75cm} (generalized $t$-distribution) \hspace{0.75cm}}}
    \startdata
    C & $0.026_{-0.019}^{+0.041}$ \\
    N & $0.070_{-0.042}^{+0.049}$ \\
    O & $0.040_{-0.028}^{+0.054}$ \\
    Na & $0.211_{-0.041}^{+0.063}$ \\
    Mg & $0.035_{-0.027}^{+0.033}$ \\
    Al & $0.068_{-0.043}^{+0.053}$ \\
    Si & $0.041_{-0.029}^{+0.036}$ \\
    S & $0.095_{-0.050}^{+0.050}$ \\
    K & $0.097_{-0.062}^{+0.067}$ \\
    Ca & $0.051_{-0.039}^{+0.071}$ \\
    Ti & $0.148_{-0.054}^{+0.052}$ \\
    V & $0.102_{-0.068}^{+0.069}$ \\
    Mn & $0.092_{-0.035}^{+0.044}$ \\
    Fe & $0.020_{-0.014}^{+0.024}$ \\
    Ni & $0.037_{-0.026}^{+0.049}$
    \enddata
    \tablecomments{\small Scatter of abundances in M67 represented by the $68^\mathrm{th}$ percentiles of hierarchical generalized $t$-distributions $f_{\nu, \hat\mu, \hat\sigma}$. Medians of the $68^\mathrm{th}$ percentiles of the distribution samplings along with their 68\% credible intervals have been reported.}
    
\end{deluxetable}
As the degrees of freedom $\nu\to\infty$, the generalized $t$-distribution approaches the normal; for smaller values of $\nu$, this distribution has heavier tails than the normal distribution, making it robust to the presence of outliers \citep{lange_1989}. We therefore use it for modeling abundance distributions.

\begin{figure}[t]
  \centering
  \includegraphics[trim=10 3 3 3, clip, width=0.8\linewidth]{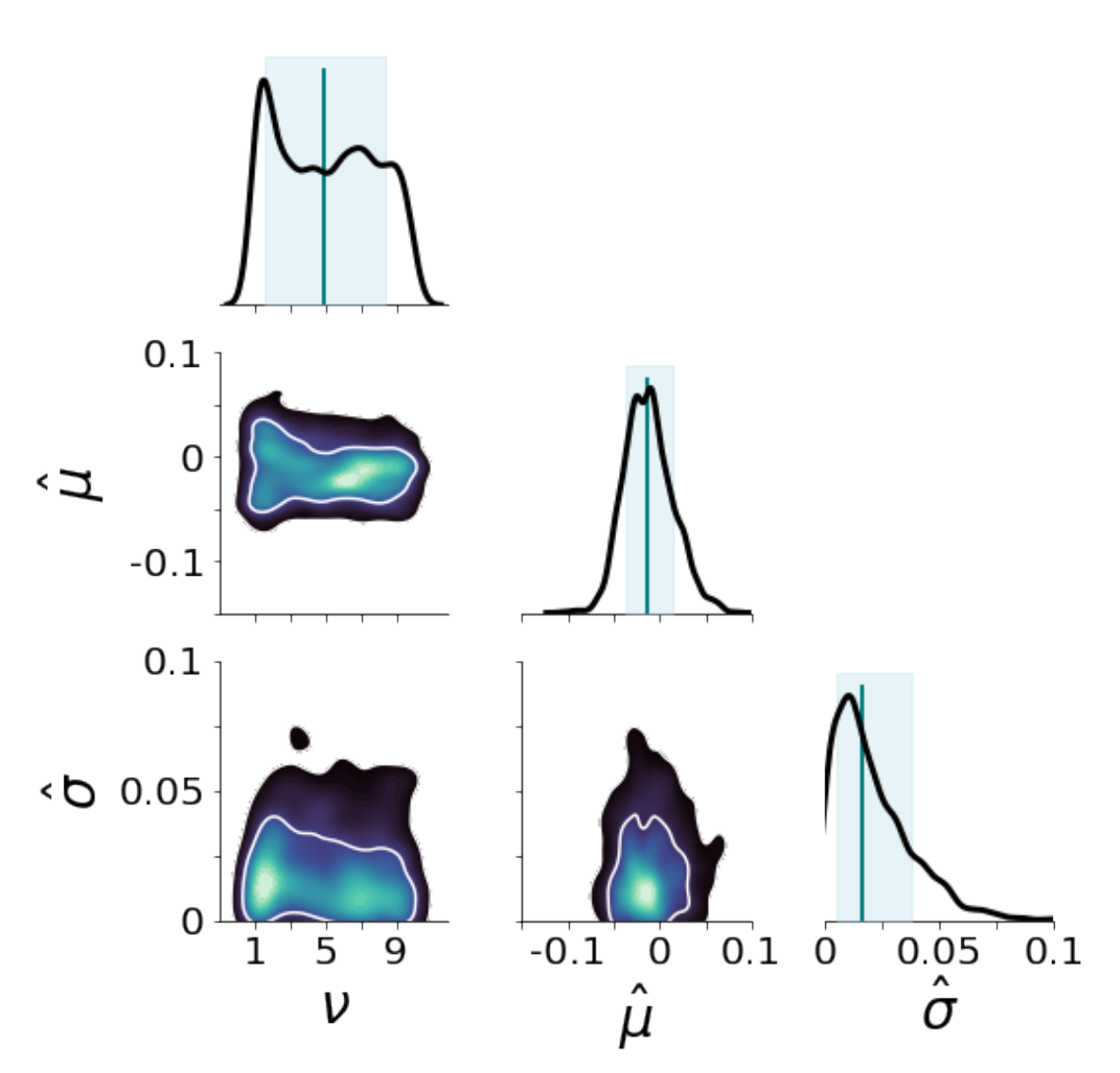}
  \caption{Parameter distributions, $\nu$, $\hat\mu$ and $\hat\sigma$, of the generalized $t$-distribution of $[\mathrm{Fe/H}]$ in M67. This is obtained through hierarchical Bayesian modeling of $[$Fe/H$]$ in M67 giant stars as described in Section \ref{subsec:LFIresults}. The inner white contour in the joint posterior PDF represents the 68\% credible interval, and the outermost boundary is at 95\%. The diagonals show the marginal distributions with their median in solid green. The median of the $68$th percentile of the $t$-distributions corresponding to these parameters is $0.020_{-0.014}^{+0.024}$, where the uncertainties encapsulate the 68\% credible interval; this is used as a proxy for $[$Fe/H$]$ scatter in M67.}
  \label{fig:FE_H_hierarch}
\end{figure}

We may evaluate whether the generalized $t$-distribution improves our hierarchical modeling beyond the normal distribution using the normality parameter $\nu$. If sufficient data are available, then it is possible to estimate $\nu$ based on the posterior; we may set $\nu=4$ for small samples, because \cite{lange_1989} show that this value is suitable for several applications. Since we have $K= 100,000$ posterior samples for $N=28$ spectra, we estimate $\nu$ using a uniform prior $\nu \sim U(1, 10)$. The lower bound of this range is set to ensure that the mean and the variance of the $t$-distributions are defined, whereas the upper bound is based on our observation that the $t$-distribution starts resembling the normal at $\nu \approx 10$.

Figure \ref{fig:FE_H_hierarch} shows the parameter distributions of the hierarchical $t$-distribution $f_{\nu, \hat\mu, \hat\sigma}([\mathrm{Fe/H}])$ in M67. We find similar distributions for other abundances.  Table \ref{tab:abundancescatter} lists the $68^\mathrm{th}$ percentiles of the hierarchical distributions for different abundances, which represent the abundance scatter in M67. Among these, $[$Fe/H$]$ shows the smallest scatter of $0.020_{-0.014}^{+0.024}$ dex, where the uncertainties on the median represent the $16$th and $84$th percentiles (or the 68\% credible interval). Most of the abundances have strong support for $\hat\sigma = 0$, that is, lack of abundance scatter in M67. However, $[\mathrm{Na/H}]$, $[\mathrm{S/H}]$, $[\mathrm{K/H}]$, $[\mathrm{Ti/H}]$, $[\mathrm{V/H}]$, and $[\mathrm{Mn/H}]$ have weaker constraints and little (but non-negligible) support for $\hat\sigma = 0$ -- this is potentially due to the weak features of Na, S, K, Ti, and V, and the scarcity of Mn lines. We discuss the reasons for the observed scatter, and whether it implies chemical homogeneity to the level of measurement precision in the next section.

\subsection{Interpretation of results}\label{subsec:LFIdiscussion}
Figure \ref{fig:M67postsingle} suggests a number of correlations between the stellar parameters and abundances. Theoretically, we expect to see many of these correlations, such as those between C, N, and O, because they are mostly determined using molecules (particularly between C and N; C and O in the plot), and those between the $\alpha$ elements, O, Mg, Si, S, Ca, and Ti (Mg and Si; O and S; O and Ca correlations are evident in the plot). However, some correlations could be due to degeneracy introduced in the statistical procedure.

For example, joint distributions of Al in Figure \ref{fig:M67postsingle} show some multi-modalities, even though Al has strong absorption features with negligible overlap with other elements; this leads to weak constraints on its marginal distribution. We also see these features in the Al marginal distributions of other M67 giants (Figure \ref{fig:abun_post}), which explains the larger than expected scatter we obtain for Al in M67 (Table \ref{tab:abundancescatter}). To understand this better, we look at the FPCs around wavelengths where Al lines are present and find that they significantly pick up the lines; this suggests that the dimensionality reduction is not a problem. We confirm this by running our procedure on mock spectra and seeing that Al is well-constrained using our 50 FPCs. Thus, the issue is likely due to a combination of the following three issues: (1) Al has very few lines that are strong and potentially exhibit nonlinear structure, (2) non-LTE effects affect Al abundances in giant and metal-poor stars, but our PSM models assume LTE \citep{nordlander_2017}, (3) the presence of noisy or missing (masked) Al lines can adversely affect dimensionally reduced data, albeit to a smaller extent due to FPCA. We could potentially resolve point (3), at least partially, by increasing the number of FPCs beyond 50; however, there is a trade-off between increasing information and decreasing the effects of noise. A similar argument can be made for Mn.

It is important to note that elements such as Na, S, K, Ca, Ti, and V, which are hard to constrain due to their weak features, could lead to a wider exploration of parameter space and some degeneracy. Figure \ref{fig:M67postsingle} shows that Ti is not well constrained, and this might be affecting constraints on other abundances.

Assuming chemical homogeneity in open clusters and taking into account observational uncertainties on spectra, we expect constraints on hierarchical distributions of $[$Fe/H$]$, $[$C/H$]$, $[$N/H$]$, and $[$O/H$]$ in M67 to be strong -- this is because they exhibit several absorption features in the near-infrared. While estimates of Fe are well constrained, those on C, N, and O face three complications. First, they are primarily constrained using their molecular features, which have complex correlated structures. Second, stellar evolutionary processes alter their surface abundances, particularly convective mixing along the giant branch (first dredge-up) and atomic diffusion \citep{souto_2019}. Third, core-collapse supernovae (CCSNe) can contaminate clusters during their formation, and C and O dominate this contamination ($\approx 0.04$ dex and $0.03$ dex increases in C and O abundances of new stars respectively due to a $60~\mathrm{M}_{\odot}$ CCSN; \citealt{bovy_2016}). In comparison, Fe contamination by CCSNe is negligible ($\approx 0.004$ dex increase due to a CCSN of the same mass; \citealt{bovy_2016}) -- stellar evolutionary effects like deep mixing or atomic diffusion do not affect Fe abundances, which are derived using several atomic absorption features.

Given that C and O distributions in M67 are constrained to $\approx 0.04$ dex suggests that pollution due to CCSNe of massive stars is likely not the case. Instead, when we look at the trends of $[$C/N$]$ with $T_\mathrm{eff}$ and $\log g$, we find evidence of mixing due to the first dredge-up as well as atomic diffusion similar to what is seen in \cite{souto_2019}. Thus, our results support their finding that stellar evolutionary effects lead to abundance variations in M67 stars and are also in line with the constraints on star formation timescales due to lack of self-pollution by CCSNe in \cite{bovy_2016}.

It is important to note that our method for estimating stellar properties differs significantly from the one adopted by APOGEE. We use PSM to model stellar spectra and infer stellar parameters and abundances simultaneously using FPCA+SNL, whereas APOGEE estimates stellar parameters through $\chi^2$ fitting of ATLAS9 (or MARCS) model atmosphere grids followed by abundance estimation using spectral windows. Therefore, estimates from the two approaches cannot be blindly compared. However, the fact that our method is capable of simultaneously estimating 17 stellar photospheric properties with negligible prior information and significant dimensionality reduction demonstrates its power.

\section{Discussion}\label{sec:discussion}
In the following sections, we discuss the advantages of our methods and their prospects for Galactic archaeology as well as astronomy in general. We also mention their limitations and highlight potential improvements.

\subsection{FPCA}
FPCA works well in the context of APOGEE stellar spectroscopy because of its ability to counter correlated uncertainties across pixels. In a typical \texttt{apStar} spectrum, multiple visits are combined after radial velocity corrections and resampling. Therefore, several noisy raw pixels contribute to a single processed pixel, which leads to correlations among pixel uncertainties. A variable line-spread function across fibers and imperfect night-sky and telluric line subtraction further exacerbate these correlations. Given such noise in the data, the registration of discrete spectral observations as smooth functions allows better estimation of the underlying processes. We can further improve this by adding domain knowledge into the procedure; using appropriate basis functions, we can capture functional dependences across wavelengths (pixels) and detect intrinsic correlations. Thus, FPCA extracts intrinsic variance from smooth covariance functions rather than blindly extracting all variance in the data \citep{viviani_2005}.

In this study, we discussed the advantages of FPCA over classical PCA/EMPCA. However, it is important to consider whether there may be a genuine signal retrieved by classical PCA/EMPCA that FPCA removes. We evaluate this by comparing the residuals of spectra reconstructed using FPCA to the expected uncertainties in APOGEE (refer to Figure \ref{fig:fpca_model}) and find no significant evidence of missing intrinsic variance. In fact, we observe that the FPCA residuals have smaller correlations across cross-sectional and longitudinal directions (across wavelengths and spectra) than those of PCA. Uncorrelated errors indicate that most of the variance that may be missing is likely due to random noise and localized systematics, which have minimal wavelength and spectral dependences.

\subsubsection{Prospects for Galactic Archaeology}
The FPCs we obtain open avenues for direct inspection and comparison of spectra to study properties of stellar populations, rather than relying on poorly constrained or biased stellar parameters and abundances \citep{bovy_2016, price_jones_2018}. As described in Section \ref{subsec:intro_chem}, mapping the chemodynamical evolution of the Milky Way galaxy through chemical tagging is a promising but challenging task given the current precision of abundance estimates. However, we can achieve this task by finding chemically similar groups of stars using the FPCs, in both the weak and strong regimes of chemical tagging. In the weak regime, we can study stellar populations to understand, e.g., the age-metallicity structure (and more generally age-abundance structures) and remnants of mergers in the Galaxy. In the strong regime, we can chemically tag stars to their birth clusters. FPCA dimensions can remove systematic trends and noise in spectra and can then be used as input to approaches like the one in \cite{mijolla_2021} to further disentangle stellar parameters from abundance information. In addition, due to its robust behavior compared to PCA in small sample studies, as illustrated in the case study of $N \ll M$, FPCA can be useful for exploratory studies of individual stellar populations. FPCA thus has the ability to revolutionize our understanding of the assembly history and chemodynamical evolution of the Galaxy.

FPCA also has promising implications for the proper uncertainty characterization of astronomical data. While astronomy has entered a golden era for Big Data applications due to large-scale, high-quality surveys, it still lacks rigorous handling of latent, nuisance variables and uncertainties. We expect that FPCA will be a step in the right direction since it not only allows the extraction of intrinsic variability, but also provides a way to explore noisy features in data. Our results indicate that ${\approx}$10 dimensions are required to characterize the stellar photospheric variance in spectra.

Our current sample of stars is not \textit{complete}, i.e., it does not include dwarfs, which can help us understand the effects of rotation and stellar evolution on abundances. In the future, we can look at the properties of dwarfs and giants simultaneously by projecting their spectra onto a functional space and employing \textit{functional data clustering} \citep{ramsay_2006} to find out whether they cluster in different subspaces. We can then apply FPCA on the clusters to reveal the differences between them.

\subsection{SNL}
The core advantage of SNL is the usage of \textit{NDEs} that learn complex distributions. State-of-the-art NDEs such as MADE and its successor, masked autoregressive flow \citep[MAF,][]{papamakarios_2017}, can learn multi-modal distributions; however, this makes them susceptible to exploring several local minima. We find that, as the networks converge, they try to locate unexpected multi-modalities in spectral likelihoods. These modalities could be due to broad, uniform priors, which allow exploration of infeasible parametric spaces or could be due to our dimensionality reduction. While this is an issue, we find that MADEs are required to some extent since MDNs alone do not track the likelihoods well. Thus, choosing and combining architectures according to the problem at hand and careful tuning and training of NDEs are necessary to ensure that the learning is accurate and robust. We can also improve the SNL inference by using informative, joint priors on the stellar parameters and abundances.

Another element of the procedure that can be significantly improved is the \textit{Simulator}. Its role is to create synthetic spectra that resemble observed data, thereby reducing the synthetic gap. Currently, we only use APOGEE uncertainty estimates and assume that they track the real noise well -- this is somewhat warranted given that FPCA removes variance due to systematics. However, the noise model can be extended to include systematics induced by (1) incompleteness of the photospheric models (e.g., non-LTE effects, three-dimensional radiative transfer, missing atomic/molecular data), (2) instrumental/extrinsic factors (e.g., non-Gaussian line-spread functions that vary across fibers, lines introduced by the interstellar medium), and (3) preprocessing of spectra using the APOGEE pipeline (e.g., radial velocity corrections, continuum normalization). In addition, we could model stellar evolutionary effects such as surface abundance variations due to convective mixing and atomic diffusion in the \textit{Simulator} to estimate birth abundances directly.

LFI is currently an active field of research in machine learning, and we expect to see significant improvements in the coming years. For example, SNRE \citep{hermans_2020, durkan_2020} is a new neural LFI method known to work well for high-dimensional data (refer to Section \ref{subsec:LFI}) but it underperforms more generally compared to LFI. We could compare our method with SNRE and see whether using a full spectrum with SNRE is better than using a dimensionally reduced spectrum with SNL.

\subsubsection{Prospects for Strong Chemical Tagging}
Our constraints on chemical homogeneity demonstrate that our SNL procedure can readily learn complex likelihoods and obtain precise and accurate stellar photospheric estimates in an efficient and robust way. We can extend this technique to other open clusters to provide tighter constraints on self-pollution by CCSNe and put limits on $\tau_\mathrm{SF}$, the time over which star-formation proceeds. Through limits on $\tau_\mathrm{SF}$, we can inform models of star cluster formation in GMCs \citep{matzner_2000}.

In order to extend our inferential method to perform fast, high-fidelity inference of the abundances of a large sample of APOGEE giant stars, we need to improve its computational time. The benefit of using simulation-based LFI is that it has the capability to perform amortized inference. An amortized posterior, $p(\theta \mid x)$, is one that spans the entire observational space, that is, it is not focused on any particular observation. We can evaluate such a posterior for different observations without having to rerun SNL each time \citep{hermans_2020}. However, the number of simulations required for the posterior to perform well across all observations scales exponentially with the diversity of observations. To tackle this, we can employ clustering techniques that find groups of \textit{similar} stellar spectra based on their FPC scores and calculate semi-amortized posteriors in spectral subspaces.  We can then efficiently infer posteriors for a large number of stars to estimate the present-day chemical distribution of the Galactic disk. This will provide opportunities for understanding the formation and evolution of the disk through strong chemical tagging.

\section{Conclusion}\label{sec:conclusion}
A key metric for success in modern spectroscopic analyses is the achievement of a high level of \textit{precision} in estimates of stellar properties. Precision is particularly important when making inferences about the formation and evolution of galactic disks based on the properties of their stellar populations. While the theory of stellar photospheres and spectroscopic instrumentation have significantly improved in the last few decades, we still face the problem of the \textit{synthetic gap} between data and models in the light of complex systematics. In this study, we use FPCA to disentangle the low-dimensional manifold of spectral data intrinsic to the stars from systematics and forgo the need to rely on model-derived stellar parameters and abundances. 

Spectroscopic data are discrete observations of continuous functions of wavelength subject to noise. We can treat them as \textit{functional data} and apply FPCA to extract the underlying functions. Through the use of basis functions, FPCA adds continuity, smoothness, and domain-specific knowledge into the computation of PCs. These features make the FPCs less vulnerable to systematics; hence, we can use them to reduce the dimensionality of spectra and impute masked regions. In contrast, classical PCA, a widely used dimensionality reduction technique in astronomy, and its extension EMPCA, are severely affected by systematics. Using a case study, we illustrate the advantages of FPCA over classical PCA and provide recommendations for its implementation from a pedagogical point of view. In particular, we help the reader make informed decisions on the type and number of basis functions ($K$) as well as the number of FPCs or dimensions ($J$) required to explain maximum intrinsic variance in functional data.

We then apply FPCA on a sample of $18,933$ giant spectra from APOGEE and find that the intrinsic spectral structure is ${\approx}10$-dimensional; this dimensionality constraint agrees with those of previous studies \citep[e.g.,][]{price_jones_2018, ting_2021}. Absorption features dominate the FPCs and they show no signs of systematics. To compare our results with previous benchmarks \citep[e.g.][]{price_jones_2018}, we perform EMPCA on the same sample and find that the EM PCs have large-scale trends attributed to persistence and other systematics. This litmus test indicates that our FPCs remove systematic features and enable direct spectral analysis for stellar and Galactic studies -- especially for chemical tagging in the Milky Way disk.

To demonstrate that our FPCs incorporate intrinsic stellar photospheric information, we use them to infer the stellar parameters ($T_\mathrm{eff}$, $\log g$) and 15 abundances of 28 giant stars in the open cluster M67 and constrain its chemical homogeneity. We apply SNL, a simulation-based LFI method, to capture non-Gaussianity in complex likelihoods such as those of real spectral data. The abundances of the M67 giants are hierarchically combined to obtain the following tight constraints on abundance scatter: $\mathrm{Fe} \lesssim 0.020_{-0.014}^{+0.02}$, $\mathrm{C} \lesssim 0.026_{-0.02}^{+0.04}$, $\mathrm{Mg} \lesssim 0.035_{-0.03}^{+0.03}$, $\mathrm{Ni} \lesssim 0.037_{-0.03}^{+0.05}$, $\mathrm{O} \lesssim 0.040_{-0.03}^{+0.05}$, $\mathrm{Si} \lesssim 0.041_{-0.03}^{+0.04}$, $\mathrm{Ca} \lesssim 0.051_{-0.04}^{+0.07}$, $\mathrm{Al} \lesssim 0.068_{-0.04}^{+0.05}$, and $\mathrm{N} \lesssim 0.070_{-0.04}^{+0.05}$, where the uncertainties indicate the 68\% credible interval. We get weaker constraints on other elemental abundances for reasons such as few or weak absorption features. Our results are in line with previous studies \citep{de_silva_2006, de_silva_2007a, de_silva_2007b, reddy_2012, ting_2012chem, bovy_2016, souto_2019} and suggest no self-pollution of M67 by core-collapse supernovae. This finding will help us understand the exact nature of star-forming clouds and their role in the formation and evolution of our Galaxy.

A significant contribution of this paper is the application and development of novel statistical methods, FPCA and SNL, in spectroscopic analyses. The fact that we obtain stringent constraints on chemical homogeneity in M67 using spectra with significant dimensionality reduction, non-Gaussian likelihoods, and uninformative priors is a testament to the power of these methods.  \textit{Functional} data sets similar to the APOGEE spectra we consider here are common in astronomy, in either the wavelength or time domain, and we envision that our results will spark wider application of FPCA in astronomy. In addition, astronomical analyses often use the assumption of Gaussian likelihoods without being warranted, and we expect that SNL and other LFI methods will provide several improvements in this regard.

\acknowledgements
A.A.P. and J.B. acknowledge financial support from NSERC (funding reference numbers RGPIN-2015-05235 \& RGPIN-2020-04712) and an Ontario Early Researcher Award (ER16-12-061). G.E. acknowledges funding from NSERC through Discovery Grant RGPIN-2020-04554 and from UofT through the Connaught New Researcher Award, both of which supported this research.

The authors thank Natalie Price-Jones for insightful conversations during the early stages of this work. We also thank Josh Speagle for providing thoughtful feedback on this paper.

Funding for the Sloan Digital Sky Survey IV has been provided by the Alfred P. Sloan Foundation, the U.S. Department of Energy Office of Science, and the Participating Institutions. SDSS-IV acknowledges support and resources from the Center for High Performance Computing  at the University of Utah. The SDSS website is \href{https://www.sdss.org/}{www.sdss.org}.

\software{\texttt{apogee} \citep{bovy_2016}, \texttt{astropy} \citep{astropy:2013, astropy:2018}, \texttt{emcee} \citep{foreman_mackey_2013}, \texttt{empca} \citep{bailey_2016}, \texttt{matplotlib} \citep{matplotlib:2007}, \texttt{numpy} \citep{numpy:2020}, \texttt{pandas} \citep{mckinney-proc-scipy-2010}, \texttt{pydelfi} \citep{alsing_2019}, \texttt{scipy} \citep{scipy:2020}, \texttt{seaborn} \citep{Waskom2021} \texttt{statsmodels} \citep{statsmodels:2010}, \texttt{tqdm} \citep{casper_da_costa_luis_2021}}.

\section*{Code availability}
The code to perform our analysis is publicly available at \href{https://github.com/aaryapatil/specdims}{https://github.com/aaryapatil/specdims}. It includes a Python implementation of the FPCA algorithm described in this paper, which can be installed and applied to different data sets.

\section*{Appendix}
Table \ref{tab:M67abundances} lists the fifteen inferred abundances of the 28 M67 giants used in this study. Each abundance estimate is the median of the corresponding marginalized distribution and the uncertainty on the estimate represents the interquartile range.
\startlongtable
\begin{longrotatetable}
\begin{deluxetable}{lRRRRRRRRRRRRRRRRRR}
\setlength{\tabcolsep}{2pt}
\tabletypesize{\tiny}
\tablecaption{\label{tab:M67abundances}Inferred Abundances of M67 Giants}
\tablewidth{0pt}
\tablehead{
\hline
\colhead{2Mass ID} &
\colhead{$[\mathrm{Fe/H}]$} &
\colhead{$[\mathrm{C/H}]$} &
\colhead{$[\mathrm{N/H}]$} &
\colhead{$[\mathrm{O/H}]$} &
\colhead{$[\mathrm{Na/H}]$} &
\colhead{$[\mathrm{Mg/H}]$} &
\colhead{$[\mathrm{Al/H}]$} &
\colhead{$[\mathrm{Si/H}]$} &
\colhead{$[\mathrm{S/H}]$} &
\colhead{$[\mathrm{K/H}]$} &
\colhead{$[\mathrm{Ca/H}]$} &
\colhead{$[\mathrm{Ti/H}]$} &
\colhead{$[\mathrm{V/H}]$} &
\colhead{$[\mathrm{Mn/H}]$} &
\colhead{$[\mathrm{Ni/H}]$}
}
\startdata
\\
2M08492491+1144057 & 0.09_{-0.11}^{+0.12} & -0.06_{-0.17}^{+0.22} & 0.17_{-0.17}^{+0.15} & -0.12_{-0.10}^{+0.11} & -0.03_{-0.19}^{+0.21} & 0.02_{-0.12}^{+0.14} & 0.30_{-0.27}^{+0.13} & 0.10_{-0.17}^{+0.15} & -0.02_{-0.23}^{+0.20} & -0.04_{-0.22}^{+0.22} & 0.09_{-0.22}^{+0.20} & 0.13_{-0.20}^{+0.17} & 0.25_{-0.32}^{+0.14} & 0.21_{-0.15}^{+0.14} & 0.01_{-0.15}^{+0.17} \\
2M08503613+1143180 & -0.00_{-0.08}^{+0.09} & -0.18_{-0.14}^{+0.13} & 0.17_{-0.18}^{+0.22} & -0.00_{-0.21}^{+0.21} & -0.07_{-0.20}^{+0.25} & 0.05_{-0.11}^{+0.10} & -0.05_{-0.20}^{+0.17} & 0.20_{-0.08}^{+0.09} & -0.04_{-0.22}^{+0.21} & -0.17_{-0.15}^{+0.25} & -0.01_{-0.25}^{+0.24} & -0.03_{-0.21}^{+0.24} & -0.00_{-0.26}^{+0.25} & 0.02_{-0.15}^{+0.16} & 0.00_{-0.16}^{+0.15} \\
2M08504964+1135089 & 0.03_{-0.05}^{+0.05} & -0.03_{-0.14}^{+0.15} & 0.11_{-0.16}^{+0.14} & -0.13_{-0.17}^{+0.19} & 0.06_{-0.22}^{+0.22} & 0.14_{-0.08}^{+0.07} & -0.08_{-0.21}^{+0.26} & 0.22_{-0.08}^{+0.07} & 0.04_{-0.19}^{+0.20} & 0.01_{-0.22}^{+0.26} & -0.08_{-0.17}^{+0.20} & -0.00_{-0.19}^{+0.22} & -0.16_{-0.18}^{+0.24} & -0.14_{-0.14}^{+0.16} & -0.10_{-0.11}^{+0.11} \\
2M08504994+1149127 & -0.02_{-0.05}^{+0.05} & -0.09_{-0.12}^{+0.15} & 0.12_{-0.17}^{+0.24} & -0.12_{-0.22}^{+0.28} & 0.08_{-0.30}^{+0.23} & -0.13_{-0.06}^{+0.06} & -0.15_{-0.18}^{+0.23} & 0.07_{-0.10}^{+0.11} & 0.05_{-0.17}^{+0.20} & -0.25_{-0.14}^{+0.20} & 0.17_{-0.19}^{+0.14} & 0.02_{-0.20}^{+0.21} & -0.05_{-0.18}^{+0.20} & -0.19_{-0.10}^{+0.10} & -0.29_{-0.08}^{+0.08} \\
2M08505816+1152223 & 0.08_{-0.11}^{+0.09} & 0.01_{-0.16}^{+0.15} & 0.26_{-0.19}^{+0.20} & -0.06_{-0.22}^{+0.21} & -0.09_{-0.18}^{+0.22} & 0.06_{-0.11}^{+0.11} & 0.07_{-0.23}^{+0.20} & 0.25_{-0.11}^{+0.10} & 0.06_{-0.26}^{+0.21} & -0.07_{-0.22}^{+0.25} & -0.06_{-0.20}^{+0.24} & 0.06_{-0.21}^{+0.21} & -0.00_{-0.23}^{+0.20} & -0.08_{-0.18}^{+0.18} & 0.02_{-0.15}^{+0.17} \\
2M08510839+1147121 & 0.07_{-0.07}^{+0.06} & -0.07_{-0.12}^{+0.15} & 0.25_{-0.26}^{+0.18} & -0.29_{-0.14}^{+0.32} & 0.04_{-0.14}^{+0.11} & 0.06_{-0.16}^{+0.16} & 0.13_{-0.28}^{+0.21} & 0.27_{-0.25}^{+0.13} & -0.04_{-0.11}^{+0.12} & -0.07_{-0.16}^{+0.13} & 0.10_{-0.13}^{+0.17} & 0.00_{-0.13}^{+0.11} & -0.06_{-0.15}^{+0.19} & -0.06_{-0.09}^{+0.07} & 0.16_{-0.14}^{+0.12} \\
2M08511269+1152423 & -0.11_{-0.12}^{+0.13} & -0.11_{-0.18}^{+0.19} & 0.17_{-0.23}^{+0.23} & 0.10_{-0.22}^{+0.18} & -0.07_{-0.23}^{+0.29} & 0.07_{-0.12}^{+0.12} & -0.08_{-0.19}^{+0.20} & 0.26_{-0.14}^{+0.11} & 0.02_{-0.18}^{+0.23} & -0.13_{-0.20}^{+0.22} & -0.01_{-0.21}^{+0.23} & 0.11_{-0.24}^{+0.19} & 0.02_{-0.21}^{+0.22} & -0.15_{-0.16}^{+0.22} & -0.09_{-0.12}^{+0.12} \\
2M08511704+1150464 & -0.01_{-0.12}^{+0.11} & -0.06_{-0.17}^{+0.09} & 0.06_{-0.43}^{+0.26} & -0.00_{-0.14}^{+0.10} & 0.15_{-0.07}^{+0.10} & 0.16_{-0.04}^{+0.04} & 0.11_{-0.09}^{+0.10} & 0.04_{-0.09}^{+0.08} & -0.05_{-0.17}^{+0.26} & 0.14_{-0.17}^{+0.16} & 0.29_{-0.08}^{+0.07} & 0.31_{-0.11}^{+0.15} & 0.43_{-0.10}^{+0.04} & 0.16_{-0.07}^{+0.07} & 0.14_{-0.22}^{+0.17} \\
2M08511710+1148160 & -0.10_{-0.06}^{+0.06} & 0.10_{-0.09}^{+0.08} & 0.26_{-0.13}^{+0.13} & 0.28_{-0.10}^{+0.09} & 0.32_{-0.06}^{+0.06} & 0.04_{-0.08}^{+0.09} & -0.11_{-0.14}^{+0.16} & 0.09_{-0.08}^{+0.10} & 0.02_{-0.16}^{+0.21} & 0.02_{-0.05}^{+0.06} & -0.14_{-0.16}^{+0.17} & -0.34_{-0.11}^{+0.19} & -0.04_{-0.16}^{+0.19} & -0.15_{-0.12}^{+0.14} & -0.12_{-0.12}^{+0.14} \\
2M08511897+1158110 & -0.04_{-0.07}^{+0.08} & -0.04_{-0.09}^{+0.07} & -0.06_{-0.12}^{+0.14} & -0.15_{-0.16}^{+0.16} & -0.26_{-0.19}^{+0.58} & 0.17_{-0.17}^{+0.13} & 0.19_{-0.05}^{+0.05} & 0.29_{-0.11}^{+0.08} & 0.21_{-0.05}^{+0.06} & 0.05_{-0.13}^{+0.22} & 0.10_{-0.18}^{+0.26} & -0.16_{-0.24}^{+0.15} & -0.24_{-0.16}^{+0.23} & 0.11_{-0.40}^{+0.20} & -0.31_{-0.05}^{+0.04} \\
2M08512156+1146061 & -0.02_{-0.16}^{+0.11} & -0.18_{-0.14}^{+0.12} & 0.50_{-0.19}^{+0.13} & -0.04_{-0.16}^{+0.16} & -0.23_{-0.19}^{+0.45} & -0.04_{-0.08}^{+0.07} & 0.01_{-0.31}^{+0.29} & 0.02_{-0.10}^{+0.11} & -0.24_{-0.13}^{+0.16} & 0.05_{-0.12}^{+0.15} & 0.37_{-0.14}^{+0.08} & -0.29_{-0.09}^{+0.11} & 0.10_{-0.18}^{+0.18} & 0.26_{-0.12}^{+0.09} & -0.30_{-0.06}^{+0.11} \\
2M08512280+1148016 & 0.00_{-0.11}^{+0.12} & 0.10_{-0.15}^{+0.14} & 0.10_{-0.19}^{+0.15} & 0.09_{-0.18}^{+0.15} & 0.16_{-0.46}^{+0.24} & 0.17_{-0.07}^{+0.05} & -0.11_{-0.16}^{+0.16} & 0.27_{-0.17}^{+0.12} & 0.24_{-0.15}^{+0.12} & 0.19_{-0.18}^{+0.11} & -0.21_{-0.15}^{+0.21} & 0.08_{-0.15}^{+0.13} & 0.16_{-0.13}^{+0.10} & 0.06_{-0.22}^{+0.28} & -0.10_{-0.13}^{+0.15} \\
2M08512618+1153520 & 0.01_{-0.07}^{+0.10} & -0.18_{-0.16}^{+0.17} & -0.04_{-0.38}^{+0.41} & -0.21_{-0.17}^{+0.17} & -0.12_{-0.15}^{+0.19} & 0.23_{-0.13}^{+0.11} & 0.21_{-0.10}^{+0.10} & 0.12_{-0.10}^{+0.12} & 0.12_{-0.16}^{+0.15} & 0.04_{-0.12}^{+0.09} & -0.01_{-0.12}^{+0.13} & 0.19_{-0.16}^{+0.16} & 0.17_{-0.11}^{+0.12} & 0.07_{-0.04}^{+0.05} & -0.16_{-0.10}^{+0.10} \\
2M08512898+1150330 & 0.03_{-0.09}^{+0.10} & -0.13_{-0.07}^{+0.08} & 0.59_{-0.21}^{+0.18} & 0.20_{-0.22}^{+0.15} & -0.01_{-0.21}^{+0.26} & -0.05_{-0.10}^{+0.11} & 0.15_{-0.20}^{+0.19} & 0.27_{-0.18}^{+0.13} & -0.15_{-0.20}^{+0.23} & -0.23_{-0.11}^{+0.11} & 0.18_{-0.11}^{+0.12} & -0.19_{-0.15}^{+0.19} & 0.13_{-0.17}^{+0.16} & -0.06_{-0.11}^{+0.12} & -0.20_{-0.15}^{+0.19} \\
2M08512935+1145275 & 0.09_{-0.07}^{+0.07} & -0.23_{-0.17}^{+0.19} & 0.55_{-0.10}^{+0.10} & 0.08_{-0.10}^{+0.10} & -0.29_{-0.12}^{+0.15} & -0.06_{-0.06}^{+0.06} & 0.03_{-0.17}^{+0.14} & 0.27_{-0.09}^{+0.08} & 0.34_{-0.14}^{+0.09} & 0.07_{-0.22}^{+0.18} & -0.04_{-0.33}^{+0.44} & -0.04_{-0.11}^{+0.14} & -0.04_{-0.19}^{+0.28} & 0.01_{-0.10}^{+0.11} & -0.09_{-0.09}^{+0.08} \\
2M08513577+1153347 & -0.04_{-0.08}^{+0.08} & 0.04_{-0.06}^{+0.06} & 0.50_{-0.05}^{+0.05} & 0.37_{-0.10}^{+0.08} & -0.38_{-0.08}^{+0.16} & -0.07_{-0.07}^{+0.07} & -0.06_{-0.08}^{+0.08} & 0.11_{-0.11}^{+0.10} & -0.18_{-0.10}^{+0.10} & 0.03_{-0.10}^{+0.10} & 0.19_{-0.17}^{+0.13} & -0.17_{-0.17}^{+0.18} & 0.22_{-0.13}^{+0.13} & -0.15_{-0.09}^{+0.08} & -0.09_{-0.08}^{+0.09} \\
2M08513862+1220141 & 0.07_{-0.09}^{+0.09} & 0.10_{-0.06}^{+0.06} & 0.73_{-0.18}^{+0.15} & 0.07_{-0.15}^{+0.15} & 0.06_{-0.23}^{+0.23} & -0.14_{-0.12}^{+0.07} & -0.12_{-0.20}^{+0.20} & -0.03_{-0.21}^{+0.21} & 0.24_{-0.25}^{+0.15} & -0.25_{-0.14}^{+0.15} & 0.36_{-0.20}^{+0.09} & 0.22_{-0.17}^{+0.14} & 0.21_{-0.27}^{+0.18} & 0.15_{-0.14}^{+0.11} & -0.16_{-0.06}^{+0.07} \\
2M08513938+1151456 & -0.07_{-0.08}^{+0.09} & -0.20_{-0.07}^{+0.08} & 0.25_{-0.05}^{+0.04} & -0.12_{-0.11}^{+0.16} & -0.37_{-0.06}^{+0.08} & 0.03_{-0.06}^{+0.09} & -0.01_{-0.22}^{+0.25} & 0.21_{-0.06}^{+0.07} & -0.11_{-0.29}^{+0.30} & 0.04_{-0.22}^{+0.18} & -0.20_{-0.12}^{+0.15} & -0.00_{-0.17}^{+0.15} & -0.14_{-0.12}^{+0.15} & 0.02_{-0.11}^{+0.13} & 0.16_{-0.08}^{+0.08} \\
2M08514235+1151230 & 0.02_{-0.05}^{+0.04} & -0.07_{-0.10}^{+0.12} & 0.27_{-0.07}^{+0.10} & -0.10_{-0.14}^{+0.17} & -0.01_{-0.18}^{+0.18} & 0.19_{-0.12}^{+0.10} & -0.32_{-0.11}^{+0.24} & 0.18_{-0.08}^{+0.08} & 0.14_{-0.08}^{+0.08} & -0.29_{-0.11}^{+0.11} & 0.03_{-0.24}^{+0.22} & -0.08_{-0.24}^{+0.27} & 0.15_{-0.15}^{+0.15} & 0.09_{-0.12}^{+0.11} & 0.12_{-0.12}^{+0.11} \\
2M08514474+1146460 & 0.01_{-0.14}^{+0.14} & 0.02_{-0.16}^{+0.15} & 0.03_{-0.16}^{+0.14} & 0.04_{-0.19}^{+0.16} & -0.03_{-0.22}^{+0.20} & 0.03_{-0.17}^{+0.16} & 0.03_{-0.22}^{+0.21} & 0.20_{-0.21}^{+0.14} & -0.18_{-0.16}^{+0.18} & 0.09_{-0.21}^{+0.21} & -0.04_{-0.16}^{+0.19} & -0.07_{-0.21}^{+0.23} & -0.12_{-0.19}^{+0.22} & -0.12_{-0.12}^{+0.12} & -0.01_{-0.11}^{+0.12} \\
2M08514507+1147459 & -0.07_{-0.10}^{+0.11} & -0.02_{-0.14}^{+0.16} & 0.13_{-0.18}^{+0.21} & -0.03_{-0.20}^{+0.19} & -0.06_{-0.21}^{+0.22} & 0.11_{-0.12}^{+0.12} & 0.00_{-0.25}^{+0.23} & 0.10_{-0.14}^{+0.13} & -0.12_{-0.20}^{+0.27} & -0.05_{-0.20}^{+0.24} & 0.03_{-0.24}^{+0.22} & -0.04_{-0.21}^{+0.21} & -0.04_{-0.22}^{+0.26} & -0.13_{-0.15}^{+0.17} & -0.19_{-0.14}^{+0.19} \\
2M08514883+1156511 & -0.17_{-0.10}^{+0.10} & 0.09_{-0.16}^{+0.11} & 0.26_{-0.12}^{+0.11} & 0.13_{-0.10}^{+0.11} & -0.21_{-0.13}^{+0.15} & -0.09_{-0.10}^{+0.08} & 0.08_{-0.22}^{+0.17} & -0.13_{-0.11}^{+0.10} & 0.13_{-0.19}^{+0.17} & 0.21_{-0.21}^{+0.16} & -0.21_{-0.16}^{+0.23} & -0.27_{-0.13}^{+0.17} & 0.05_{-0.28}^{+0.23} & 0.04_{-0.13}^{+0.13} & -0.15_{-0.16}^{+0.14} \\
2M08515952+1155049 & -0.03_{-0.04}^{+0.06} & 0.02_{-0.15}^{+0.17} & 0.05_{-0.22}^{+0.18} & 0.04_{-0.24}^{+0.18} & 0.06_{-0.14}^{+0.13} & 0.19_{-0.17}^{+0.09} & -0.34_{-0.10}^{+0.14} & 0.16_{-0.04}^{+0.07} & 0.01_{-0.14}^{+0.11} & -0.15_{-0.17}^{+0.18} & -0.00_{-0.22}^{+0.17} & -0.33_{-0.10}^{+0.17} & -0.23_{-0.13}^{+0.14} & 0.13_{-0.06}^{+0.11} & -0.14_{-0.07}^{+0.12} \\
2M08521097+1131491 & -0.10_{-0.12}^{+0.10} & -0.02_{-0.17}^{+0.19} & 0.39_{-0.13}^{+0.16} & -0.08_{-0.19}^{+0.32} & 0.09_{-0.20}^{+0.19} & 0.07_{-0.07}^{+0.09} & 0.11_{-0.16}^{+0.14} & 0.14_{-0.19}^{+0.18} & -0.04_{-0.17}^{+0.17} & -0.12_{-0.18}^{+0.27} & -0.01_{-0.18}^{+0.18} & 0.11_{-0.22}^{+0.20} & -0.11_{-0.21}^{+0.26} & 0.26_{-0.07}^{+0.07} & -0.02_{-0.13}^{+0.13} \\
2M08521656+1119380 & -0.05_{-0.07}^{+0.06} & 0.05_{-0.13}^{+0.15} & 0.46_{-0.19}^{+0.17} & 0.28_{-0.11}^{+0.11} & -0.32_{-0.13}^{+0.08} & 0.17_{-0.13}^{+0.12} & 0.21_{-0.12}^{+0.11} & 0.04_{-0.09}^{+0.11} & -0.06_{-0.20}^{+0.20} & -0.00_{-0.07}^{+0.13} & -0.05_{-0.26}^{+0.18} & 0.20_{-0.04}^{+0.05} & -0.09_{-0.08}^{+0.10} & 0.12_{-0.17}^{+0.15} & -0.01_{-0.06}^{+0.08} \\
2M08522003+1127362 & 0.06_{-0.06}^{+0.04} & -0.21_{-0.10}^{+0.09} & 0.69_{-0.21}^{+0.15} & 0.14_{-0.38}^{+0.16} & 0.43_{-0.09}^{+0.04} & 0.09_{-0.04}^{+0.04} & 0.23_{-0.08}^{+0.08} & -0.08_{-0.06}^{+0.06} & 0.08_{-0.17}^{+0.26} & -0.28_{-0.13}^{+0.37} & -0.06_{-0.19}^{+0.15} & 0.01_{-0.14}^{+0.13} & -0.06_{-0.23}^{+0.21} & -0.10_{-0.08}^{+0.08} & -0.02_{-0.07}^{+0.06} \\
2M08522636+1141277 & -0.10_{-0.06}^{+0.06} & -0.06_{-0.10}^{+0.10} & 0.19_{-0.21}^{+0.20} & 0.12_{-0.16}^{+0.15} & 0.35_{-0.12}^{+0.09} & 0.00_{-0.04}^{+0.05} & 0.08_{-0.08}^{+0.09} & -0.12_{-0.06}^{+0.07} & 0.36_{-0.15}^{+0.09} & 0.33_{-0.07}^{+0.06} & -0.01_{-0.10}^{+0.09} & 0.41_{-0.07}^{+0.05} & -0.37_{-0.08}^{+0.12} & -0.27_{-0.07}^{+0.08} & -0.16_{-0.06}^{+0.06} \\
2M08525625+1148539 & -0.03_{-0.15}^{+0.13} & -0.02_{-0.09}^{+0.10} & 0.15_{-0.22}^{+0.18} & -0.03_{-0.11}^{+0.16} & -0.25_{-0.14}^{+0.16} & 0.11_{-0.05}^{+0.06} & 0.18_{-0.17}^{+0.15} & -0.06_{-0.14}^{+0.13} & -0.12_{-0.22}^{+0.25} & -0.31_{-0.10}^{+0.12} & 0.16_{-0.27}^{+0.20} & 0.13_{-0.18}^{+0.17} & 0.05_{-0.21}^{+0.19} & 0.19_{-0.16}^{+0.14} & -0.03_{-0.18}^{+0.17} \\
\\
\enddata
\tablewidth{0pt}
\tablecomments{Medians of the marginalized distributions are displayed along with their uncertainties based on the interquartile range.}
\end{deluxetable}
\end{longrotatetable}


\end{document}